\documentclass[aos,preprint]{imsart}
\RequirePackage[OT1]{fontenc}
\RequirePackage{amsthm,amsmath}

\RequirePackage[colorlinks,citecolor=blue,urlcolor=blue]{hyperref}

\usepackage{graphicx}
\usepackage{amsmath}
\usepackage{graphics}
\usepackage{epsfig}
\usepackage{amssymb}
\usepackage{color}

\usepackage[authoryear]{natbib}
\newcommand{\T}{\mathcal{T}}

\newcommand{\be}{\begin{eqnarray}}
\newcommand{\ee}{\end{eqnarray}}
\newcommand{\bea}{\begin{eqnarray*}}
\newcommand{\eea}{\end{eqnarray*}}

\newcommand{\BLUE}{\scalebox{0.5}{$\mathrm{BLUE}$}}

\newcommand{\tsum}{{\textstyle\sum}}
\newcommand{\tprod}{{\textstyle\prod}}

\newtheorem{theorem}{Theorem}[section]
\newtheorem{lemma}{Lemma}[section]
\newtheorem{proposition}{Proposition}[section]
\newtheorem{corollary}{Corollary}[section]

\newtheorem{example}{Example}[section]

\newtheorem{remark}{Remark}[section]


\begin{document}

\begin{frontmatter}
\title{Best linear unbiased estimators in continuous time regression models}
\runtitle{BLUE  in continuous time regression models}

\begin{aug}
\author{\fnms{Holger} \snm{Dette}\thanksref{m1}\ead[label=e1]{holger.dette@rub.de}},
\author{\fnms{Andrey} \snm{Pepelyshev}\thanksref{m2}\ead[label=e2]{pepelyshevan@cf.ac.uk}}
\and
\author{\fnms{Anatoly} \snm{Zhigljavsky}\thanksref{m2}\ead[label=e3]{zhigljavskyaa@cf.ac.uk}
}
\runauthor{H. Dette et al.}

\affiliation{Ruhr-Universit\"at Bochum\thanksmark{m1}  and Cardiff  University\thanksmark{m2}}

\address{Ruhr-Universit\"at Bochum \\
Fakult\"at f\"ur Mathematik \\
44780 Bochum \\
Germany\\
\printead{e1}}

\address{School of Mathematics\\
 Cardiff  University\\
  Cardiff, CF24 4AG\\
  UK\\
\printead{e2}\\
\printead{e3}}
\end{aug}

\begin{abstract}
In this paper  the problem of best linear unbiased estimation is investigated for continuous-time regression models.  We prove several general statements concerning the explicit form of the best linear unbiased estimator
(BLUE), in particular
when the error process is a smooth process
with one or several derivatives of the response process available for construction of  the estimators.
We derive the explicit form of the BLUE for many specific models  including the cases of continuous
 autoregressive errors of order two and integrated error processes (such as integrated Brownian motion).
 The results are illustrated by several  examples.
\end{abstract}



\end{frontmatter}


\section{Introduction}
\def\theequation{1.\arabic{equation}}
\setcounter{equation}{0}

Consider a  continuous-time   linear regression model of the form
\be
 y(t)=\theta^Tf(t)+\epsilon(t)\, , \quad t \in [A,B],
 \label{eq:model}
\ee
where $\theta \in \mathbb{R}^m$ is a vector of unknown parameters, $f(t)=(f_1(t), \ldots, f_m(t))^T$ is
a vector  of linearly independent functions defined
on some interval, say $[A,B]$, and $\epsilon = \{ \epsilon(t) | t \in [A,B] \}$ is
a random error process with $\mathbb{E}[\epsilon(t)]=0$ for all $t \in [A,B]$ and
covariances  $\mathbb{E}[\epsilon(t) \epsilon(s)] =K(t,s)$.
We will assume that~$\epsilon$ has continuous (in the mean-square sense) derivatives $\epsilon^{(i)} $ $(i=0,1, \ldots, q)$ up to order $q$,
where $q$ is a non-negative integer.

This paper is devoted to studying  the best linear unbiased estimator (BLUE) of the parameter $\theta$ in the general setting
and in many specific instances.
Understanding of the explicit form of the BLUE has profound significance on general estimation theory and
on asymptotically optimal design for (at least) three reasons.
{Firstly,
the   efficiency of the ordinary least squares estimator, the discrete BLUE and other unbiased estimators
 can be computed exactly. Secondly, as pointed out in a series of papers  \cite{SY1966,SY1968,SY1970},
the explicit form of the BLUE is the key ingredient  for  constructing  the (asymptotically)  optimal exact designs in the regression model
\be
 y(t_i)=\theta^Tf(t_i)+\epsilon(t_i)\, , \quad A \leq t_1  <  t_2 \ldots < t_{N-1} < t_N \leq B\,,
 \label{eq:modeldisc}
\ee
with $\mathbb{E}[\epsilon(t_i) \epsilon(t_j)] =K(t_i,t_j)$. Thirdly, simple and very efficient estimators  for the parameter $\theta$ in
the regression model \eqref{eq:modeldisc} can be derived from the continuous BLUE, like the extended signed least
squares estimator investigated in \cite{DetPZ2016} and the   estimators based on approximation of stochastic
integrals  proposed in  \cite{DetKonZ2016}.}

There are many classical papers dealing with construction of the BLUE,
mainly in the case of a non-differentiable error process; that is,  in  model \eqref{eq:model} with $q=0$.
{In this situation, it is well understood that
solving specific instances of an equation of Wiener-Hopf type
\be
\int_A^B K (t,s) \zeta  (dt)= f(s),
 \label{eq:suff-ness-cond-q0}
\ee
for an $m$-dimensional vector $ \zeta  $  of signed measures  implies an explicit construction of the BLUE
in the continuous-time model \eqref{eq:model}.
This equation was first considered in a seminal paper of
\cite{grenander1950} for the  case of the location-scale model $y(t)=\theta+\epsilon(t)$, i.e. $m=1$, $f_1(t)=1$.
   For a general regression  model with $m\geq 1$ regression functions (and $q=0$), the BLUE was extensively discussed in
 \cite{grenander1954estimation}  and   \cite{rosenblatt1956some} who considered  stationary processes in
discrete time, where the spectral representation of the error process was heavily used for the construction of the estimators.
In this and many other papers including
\cite{PisRoz1963,kholevo1969estimates,hannan1975linear} the subject of the study was concentrated around
the spectral representation of the estimators and
hence the  results in these references are only  applicable to very specific models.}
A more direct investigation of  the BLUE in the location scale model (with $q=0$) can be found in
\cite{Hajek1956}, where equation \eqref{eq:suff-ness-cond-q0} for the BLUE was solved for a few simple kernels.
The most influential paper on properties of continuous BLUE and its relation to the reproducing kernel Hilbert spaces (RKHS) is \cite{parzen1961approach}. 
A relation between discrete and continuous BLUE has been further addressed in \cite{anderson1970efficient}.
 An excellent survey of classical results on the BLUE is given in the book of \cite{N1985a}, Sect. 2.3 and Chapter 4 (for the location scale model).
Formally, Theorem 2.3  of \cite{N1985a} includes the case when the derivatives of the process $y(t)$ are available ($q\geq 0$);
this is made possible by the use of  generalized functions which may contain derivatives of the Dirac delta-function.
This theorem, however, provides only a sufficient condition for an estimator to be the BLUE.

{The main reason why results on the BLUE in model \eqref{eq:model} are so difficult to obtain consists in the fact that - except in
the location-scale model  - the functional to be minimized is  not convex, so that  the usual arguments are not applicable.}
The main  examples, where  the explicit form of the BLUE was known before the publication of the monograph by \cite{N1985a},
are listed in Sect. 2.3 of this book. In most of  these examples either a Markovian structure of the error process is assumed or
 one-dimensional location scale model is studied.
Section~\ref{sec:no-deriv} of our paper updates this list and gives a short outline of previously known cases
where the explicit form of the BLUE was known until now.

There was also an extensive study of the relation between solutions of the Wiener-Hopf equations and the BLUE
through the RKHS theory,
see  \cite{parzen1961approach,SY1966,SY1968,SY1970} for an early or \cite{ritter2000average} for a more recent reference.
If $q=0$ then the main  RKHS  assumption is usually formulated as  the existence of a solution, say $\zeta_0$, of equation \eqref{eq:suff-ness-cond-q0},
where the measure $\zeta_0$ is continuous and has {no} atoms, see \cite{berlinet2011reproducing} for the  theory of the RKHS.
As shown in the present paper, this almost never happens for the commonly used covariance kernels
and regression functions (a single general  exception from this observation  is given in Proposition~\ref{th:KL}).
Note also that the numerical construction of the continuous BLUE is difficult even for $q=0$ and $ m=1$, see e.g. \cite{ramm1980theory}
and a remark on p.80 in \cite{SY1966}. For  $q>0$, the problem  of numerical construction of the BLUE is severely ill-posed and hence is extremely hard.

{The main purpose of this paper is to provide   further insights into the structure of the BLUE (and its covariance matrix) from the observations $\{ Y(t) | t \in \T
 \}$ (and its $q$ derivatives)
in continuous-time regression models of the form \eqref{eq:model}, where the set  $ \T \subseteq [A,B]$ defines the region where the process is observed. 
By generalizing the celebrated Gauss-Markov theorem, we derive new characterizations for  the BLUE
which can be used to  determine its  explicit form and the corresponding covariance matrix in numerous models.
In particular, we do not have to restrict ourselves to one-dimensional regression models and to Markovian error processes.
Thus our results require minimal assumptions regarding the regression function  and  the error process.
Important new examples, where the BLUE can be determined explicitly, include the process with triangular covariance function \eqref{eq:triag_kernel}, 
general integrated processes (in particular, integrated Brownian motion) and continuous autoregressive   processes including the
Mat\'{e}rn kernels with parameters $3/2$ and $5/2$.  

The remaining part of this paper is organized as follows.
In Section \ref{BLUE} we  develop a consistent general theory of best linear unbiased estimation using   signed matrix measures. We are able to
circumvent the convexity problem
and derive  several important characterizations and  properties of the BLUE.
{In particular,
in Theorem~\ref{th:GenGMq1}  we provide necessary and sufficient conditions for an estimator to be BLUE when $q \geq 0$;
 in Theorem~\ref{th:GenGMq4}  such  conditions are derived for $q = 0$,  $\T \subset \mathbb{R}^d$ with $d \geq 1$ and very general assumptions about
the vector of regression functions $f(\cdot)$ and the covariance kernel $K(\cdot, \cdot)$.
}

Section \ref{sec:BLUE-C1} is devoted to models,
where the error process has  one derivative.
In particular, we derive   an  explicit form of the BLUE, see Theorems~\ref{th:BLUE-2} and \ref{th:integr}, and obtain
the BLUE for specific types of smooth kernels. In Section  \ref{sec:AR2} we consider regression models with a continuous-time autoregressive (AR)
error process of order
$2$ (i.e. CAR($2$)) in more detail. Moreover, in an online supplement [see \cite{supplement}] we demonstrate that the covariance matrix of
the  BLUE in this model can be obtained as a limit of the covariance matrices
of  the BLUE in  discrete  regression models \eqref{eq:modeldisc} with observations at equidistant points and a discrete AR($2$) error process.
In Section \ref{sec:ARp+IBM} we give some   insight into the structure of the BLUE  when the error process is more than once differentiable.
Some numerical illustrations are given in Section \ref{sec:numer}, while
 technical proofs  can be found in Section \ref{sec:appendix}.
}

\section{General linear estimators and the BLUE}
\label{BLUE}
\def\theequation{2.\arabic{equation}}
\setcounter{equation}{0}

\subsection{Linear estimators and their properties}

Consider the regression model \eqref{eq:model} with covariance kernel  $K(t,s)=\mathbb{E}[\epsilon(t) \epsilon(s)] $.
Suppose that we can observe the process $\{ y(t) | t \in \T \} $ along with its $q \geq 0$
mean square  derivatives $\{ y^{(i)}(t)| t \in \T \} $ for $i=1,\ldots,q$,
where the design set $\T$ is a Borel subset of some interval $[A,B]$ with $-\infty \leq A <B \leq \infty$.
This is possible when the kernel $K(t,s)$ is $q $ times continuously differentiable on the square $[A,B] \times [A,B]$
and the vector-function $f(t)=(f_1(t), \ldots, f_m(t))^T$  is   $q$ times  differentiable
on the interval $[A,B]$ with derivatives $f^{(1)}, \ldots f^{(q)} $
$(f^{(0)}= f)$.
Throughout this paper we will also assume that   the functions $f_1, \ldots, f_m$ are linearly independent on $\T$.

Let $Y(t)= \{(y^{(0)}(t), \ldots, y^{(q)}(t))^T \}$ be the observation vector containing the process $y(t)=y^{(0)}(t)$ and its $q$ derivatives.
Denote by $\mathbf{Y}_\T= \{ Y(t):\; t \in \T \}$ the set of all available observations.
The  general linear estimator of the parameter $\theta$ in the regression model \eqref{eq:model} can be defined as
\be
\label{eq:est}
 \hat\theta_{G}= \int_\T G(dt)Y(t) = \sum_{i=0}^q \int_\T y^{(i)}(t)G_i(dt),
\ee
where $G(dt)= (G_0(dt),\ldots,G_q(dt))$ is a matrix of size $m \times (q+1)$.
The columns of this matrix are  signed vector-measures $G_0(dt),\ldots,G_q(dt)$
defined on Borel subsets of  $\T$ (all vector-measures in this paper are signed and have length $m$).

The following lemma shows a simple way of constructing unbiased estimators;
this lemma will also be used for deriving the BLUE in many examples.
The proof is given in Section \ref{sec:appendix}.

\begin{lemma}
\label{lem:constr-unbiased}
Let $\zeta_0,\ldots,\zeta_q$
be some signed vector-measures  defined on $\T$ such that the $m\times m$ matrix
\be
\label{eq:C}
C= \sum_{i=0}^q \int_\T \zeta_i(dt) \left(f^{(i)}(t)\right)^T
\ee
is non-degenerate. Define $G=(G_0,\ldots,G_q)$,
where $G_i$ are the signed vector-measures and $G_i(dt)=C^{-1} \zeta_i(dt)$ for $i=0, \ldots, q$. Then the estimator
$\hat\theta_{G}$ is unbiased.
\end{lemma}

The covariance matrix of any unbiased estimator $\hat\theta_{G}$ of the form  \eqref{eq:est} is given by
\be
\label{eq:var}
 \mathrm{Var}(\hat\theta_{G})&=&\int_\T \int_\T  G(dt) \mathbf{K}(t,s) G^T(ds)\\
  &=&
 \sum_{i=0}^q \sum_{j=0}^q \int_\T \int_\T \frac{\partial^{i+j}K(t,s)}{\partial t^i\partial s^j}G_i(dt)G_j^T(ds)\, \nonumber,
\ee
where
$$
\mathbf{K}(t,s)= \left( \frac{\partial^{i+j}K(t,s)}{\partial t^i\partial s^j} \right)_{i,j=0}^q  ~=~\left( \mathbb{E}[\epsilon^{(i)}(t) \epsilon^{(j)}(s)]
 \right)_{i,j=0}^q  \,
$$
is the matrix consisting of the derivatives of $K$.

\subsection{The BLUE}

If there exists a set of signed vector-measures, say $G=(G_0,\ldots,G_q)$, such that the estimator
 $\hat\theta_{G}=\int_\T G(dt)Y(t)$ is unbiased
and
$
 \mathrm{Var}(\hat\theta_{H})\ge\mathrm{Var}(\hat\theta_{G}),
$
where  $ \hat\theta_{H}= \int_\T H(dt)Y(t)$ is  any other linear unbiased estimator which uses the observations $\mathbf{Y}_\T$,
then $\hat\theta_{G}$ is called the best linear unbiased estimator (BLUE) for the regression model \eqref{eq:model}
using the set of observations $\mathbf{Y}_\T$.
The BLUE depends on the kernel  $K$, the vector-function $f$, the
set $\T$ where the observations are taken and on the number $q$ of available derivatives of the process $\{y(t)|t \in \T\}$.


The following theorem is a generalization of the celebrated Gauss-Markov theorem
(which is usually formulated for the case when  $q=0$ and $\T$ is finite)  and gives a necessary and sufficient condition for an estimator to be the BLUE.
In this theorem and  below
we denote  the partial derivatives of the kernel $K(t,s)$ with respect to the first component by
$$
 K^{(i)}(t,s)=\frac{\partial^{i}K(t,s)}{\partial t^{i}}\, .
$$
The proof of the theorem can be found in  Section \ref{sec:appendix}.

\begin{theorem}
\label{th:GenGMq1}
Consider the regression model \eqref{eq:model}, where the error process $\{\epsilon(t)|t \in [A,B]\}$ has a covariance kernel $K(\cdot,\cdot) \in C^q([A,B]\times[A,B])$
and $f(\cdot) \in C^q([A,B])$ for some $q \geq 0$.
Suppose that the process $\{y(t)| t \in [A,B]\}$ along with its $q $ derivatives can be observed at all $t\in \T \subseteq [A,B]$.

An unbiased  estimator
$\hat\theta_{G} = \int_\T G(dt)Y(t)$  is  BLUE if and only if
the equality
\be
  \sum_{i=0}^q \int_\T  K^{(i)}(t,s)G_i(dt)= D f(s),
 \label{eq:suff-ness-cond-q}
\ee
is fulfilled for  all $s\in \T$, where $D$ is some $m \times m$ matrix. In this case,
$
D=\mathrm{Var}(\hat\theta_{G})
$
with $\mathrm{Var}(\hat\theta_{G})$  defined in \eqref{eq:var}.
\end{theorem}

The next proposition  is slightly weaker than    Theorem~\ref{th:GenGMq1} (here
the covariance matrix of the BLUE is assumed to be non-degenerate)
but   will  be  very useful
in further considerations.

\begin{proposition}
\label{th:GenGMq} Let the assumptions of Theorem~\ref{th:GenGMq1} be satisfied and let
  $\zeta_0,\ldots,\zeta_q$
be  signed vector-measures  defined on $\T$ such that the  matrix $C$ defined in
\eqref{eq:C}
is non-degenerate. Define
 $G=(G_0,\ldots,G_q)$,  $\;G_i(dt)=C^{-1} \zeta_i(dt)$ for $i=0, \ldots, q$.
The estimator
 $\hat\theta_{G} = \int_\T G(dt)Y(t)$ is the BLUE if and only if
 \be
  \sum_{i=0}^q \int_\T  K^{(i)}(t,s)\zeta_i(dt)= f(s)
 \label{eq:suff-cond-q}
\ee
 for  all $s\in \T$.
In this case, the covariance matrix of  $\hat\theta_{G}$ is $\mathrm{Var}(\hat\theta_{G})=C^{-1}$.
\end{proposition}

\subsection{Grenander's theorem and its generalizations}$\,$

\label{sec:Grenander}

When $\T=[A,B]$, $q=0$, $m=1$ and the regression model \eqref{eq:model} is the location-scale model $y(t)=\alpha + \varepsilon(t)$, Theorem~\ref{th:GenGMq1}
is known as Grenander's theorem [see \cite{grenander1950} and Section 4.3 in \cite{N1985a}].
In this special case
Grenander's theorem has been generalised by  \cite{N1985a}
to the case when $\T\subset \mathbb{R}^d$ [see Theorem 4.3 in this reference].
The reason why  Grenander's and N{\"{a}}ther's theorems only deal with the location-scale model
is caused by  the fact that for this model   the convexity of the functional to be minimized is easy to establish. For general regression models
the convexity argument is  not directly applicable and hence the problem is much harder.

For the case of one-dimensional processes,  Theorem~\ref{th:GenGMq1} generalizes Grenander's theorem to arbitrary $m$-parameter regression models of the form
 \eqref{eq:model} and the case of arbitrary $q \geq 0$.
Another generalization of the  Grenander's theorem is given below; it deals with a general $m$-parameter regression model \eqref{eq:model}
with a continuous error process (i.e. $q=0$) and
a $d$-dimensional  set $\T \subset \mathbb{R}^d$; that is, the case where $y(t)$ is a random field.
Note that the conditions on the vector of regression functions $f(\cdot)$
in Theorem~\ref{th:GenGMq4} are weaker (when $d=1$) than
the conditions on $f(\cdot)$
in Theorem~\ref{th:GenGMq1}  applied in the case $q=0$.

\begin{theorem}
\label{th:GenGMq4}
Consider the regression model $y(t)=\theta^T f(t)+\epsilon(t) $, where $ t \in \T \subset \mathbb{R}^d$,
the error process $\epsilon(t)$ has  covariance kernel $K(\cdot,\cdot)$
and \mbox{$f\!:  \T \!\to \! \mathbb{R}^m $} is a vector of bounded integrable and linearly independent functions.
Suppose that the process $y(t)$ can be observed at all $t\in \T$ and let $G$ be a signed vector-measure on $\T$, such that the
  estimator
$
\hat\theta_{G} = \int_\T G(dt)Y(t)
$    is unbiased. $\hat \theta_G$ is a  BLUE if and only if
the equality
$$
  \int_\T  K(t,s)G(dt)= D f(s)
$$
holds for  all $s\in \T$ for  some $m \times m$ matrix $D$. In this case,
$
D=\mathrm{Var}(\hat\theta_{G}),
$
where $\mathrm{Var}(\hat\theta_{G})$
 is the covariance matrix of $\hat\theta_{G}$
defined by \eqref{eq:var}.

\end{theorem}

The proof of this theorem is a simple extension of the proof of Theorem~\ref{th:GenGMq1}
with $q=0$ to general $\T \subset \mathbb{R}^d$ and left to the reader.

\subsection{Properties of the BLUE}

\begin{itemize}
  \item[(P1)]  Let $\hat\theta_{G_1}$ and $\hat\theta_{G_2}$ be  BLUEs for the same regression model \eqref{eq:model} and the same $q$  but for two different design sets  $\T_1 $ and $\T_2$ such that $\T_1 \subseteq \T_2$. Then $\mathrm{Var}(\hat\theta_{G_1}) \geq \mathrm{Var}(\hat\theta_{G_2})$.
  \item[(P2)]
  Let $\hat\theta_{G_1}$ and $\hat\theta_{G_2}$ be  BLUEs for the same regression model \eqref{eq:model} and the same design set $\T$  but for two different values of $q$, say, $q_1 $ and $ q_2$, where $0\leq q_1 \leq q_2$. Then $\mathrm{Var}(\hat\theta_{G_1}) \geq \mathrm{Var}(\hat\theta_{G_2})$.
\item[(P3)]
  Let $\hat\theta_{G}$ with  $G=(G_0,\ldots,G_q)$ be a BLUE for the regression model \eqref{eq:model},   design space $\T$  and given $q\geq 0$.
  Define $g(t)=L f(t)$, where $L$ is a non-degenerate $m \times m $ matrix, and a signed vector-measure
   $H=(H_0,\ldots,H_q)$ with $H_i(dt)=L^{-1} G_i(dt)$ for $i=0,\ldots, q$. Then $\hat\theta_{H}$ is a BLUE for
     the regression model $y(t)= \beta^T g(t)+ \varepsilon(t)$   with the same $y(t)$, $\varepsilon(t)$, $\T$  and~$q$. The covariance matrix of
     $\hat\theta_{H}$ is $L^{-1}\mathrm{Var}(\hat\theta_G){L^{-1}}^T$.

\item[(P4)]
 If $\T=[A,B]$ and a BLUE  $\hat\theta_{G}$ is defined by
 the matrix-measure $G$ that has smooth enough continuous parts, then  we can choose another representation  $\hat\theta_{H}$  of the same  BLUE, which is defined by
  the matrix-measure $H=(H_0,H_1,\ldots,H_q)$ with
 vector-measures $H_1, \ldots, H_q$ having no continuous parts.

\item[(P5)]
Let   $\zeta_0,\ldots,\zeta_q$ satisfy the equation \eqref{eq:suff-cond-q}
for all $s \in \T$, for some vector-function $f(\cdot)$,
  design set $\T$  and given $q\geq 0$. Define $C=C_f$ by \eqref{eq:C}.
Let $g(\cdot)$ be some other $q$ times differentiable vector-function on the interval $[A,B]$.
Assume that for all $s \in \T$, signed vector-measures  $\eta_0,\ldots,\eta_q$ satisfy the equation
 \be \label{eq:g}
  \sum_{i=0}^q \int_\T  K^{(i)}(t,s)\eta_i(dt)= g(s);
\ee that is, the equation \eqref{eq:suff-cond-q}
 for the  vector-function $g(\cdot)$, the same design set $\T$  and the same  $q$. Define  $C_g=\sum_{i=0}^q \int_\T g^{(i)} (t) \eta_i^T(dt)
$, which is  the matrix \eqref{eq:C}
 with $\eta_i$ substituted for $\zeta_i$ and $g(\cdot)$ substituted for $f(\cdot)$.

If the matrix $C=C_f+C_g$ is non-degenerate,  then we define
 the set of signed vector-measures   $G=(G_0,\ldots,G_q)$  by
 $G_i=C^{-1}(\zeta_i+\eta_i),$ $i=0,\ldots, q$, yielding  the estimator $\hat\theta_{G} $. This estimator is a   BLUE
  for the regression model $y(t)=\theta^T[f(t)+g(t)]+\varepsilon(t)$, $t \in \T$. 
  \end{itemize}

Properties (P1)--(P3) are obvious. The property (P4) is a particular case of the discussion of Section~\ref{sec:non_uniq}. To prove (P5)
we simply add the equations \eqref{eq:suff-cond-q} and \eqref{eq:g} and then use Proposition~\ref{th:GenGMq}.

We believe that the properties (P4) and (P5) have never been noticed before and both these properties
are very important for understanding best linear unbiased estimators in the continuous-time regression model
\eqref{eq:model}  and especially for
constructing a BLUE for new models from the cases when a  BLUE is known for simpler models.
As an example, assume that all functions in the vector $f$ are not constant and set $g(t)=c$, where $c$ is a constant vector.
Then, if we know the BLUE for $f$ and another BLUE for the location-scale model,
 we can use  property (P5) to
construct BLUE for $\theta^T(f(t)+c)$. This is an essential part of the proof
of Theorem~\ref{th:integr}, which allows obtaining the explicit form of the BLUE for
the integrated error processes from the explicit form of the BLUE for the corresponding non-integrated errors (which is a much easier problem).

\subsection{Non-uniqueness}
\label{sec:non_uniq}
Let us show  that if $\T=[A,B]$ then, under the additional smoothness conditions,
for a given set of signed vector-measures  $G=(G_0,G_1, \ldots, G_q)$ on $\T$ we can  find another set of measures
 $H=(H_0,H_1, \ldots, H_q)$ such that the signed
 vector-measures $H_1, \ldots, H_q$ have no continuous parts but the expectations and covariance matrices of the
 estimators $\hat\theta_{G}$ and $\hat\theta_{H}$ coincide.

For this purpose, let $q>0$, $G_0,\ldots,G_q$ be some signed vector-measures and for some $i \in \{1, \ldots, m\}$,
the signed measure $G_i(dt)$ has the form
$$G_i(dt)=Q_i(dt)+{\varphi}(t)dt,$$
 where $Q_i(dt)$ is a signed vector-measure and ${\varphi} \in C^i([A,B])$
 (that is, $\varphi$ is an $i$ times differentiable vector-function on the interval $[A,B]$).
Define the matrix $H=(H_0,\ldots,H_q)$, where the columns of $H$ are the following signed vector-measures:
\bea
H_0(dt)=G_0(dt)+(-1)^i \left[{\varphi}^{(i)}(t)dt-\varphi^{(i-1)}(A)\delta_A(dt)+\varphi^{(i-1)}(B)\delta_B(dt)\right],
\eea
$H_i(dt)= Q_i(dt),$ $
  H_j(dt)= G_j(dt),$ for $j=i+1,\ldots,q$ and
  \bea
 H_j(dt)= G_j(dt)+(-1)^{i-j-1} \left[\varphi^{(i-j-1)}(A)\delta_A(dt)-\varphi^{(i-j-1)}(B)\delta_B(dt)\right]\;\;\;\;\;\;
\eea
 for $j=1,\ldots,i-1,$ where
$\delta_A(dt)$ and $\delta_B(dt)$ are the Dirac delta-measures concentrated at the points $A$ and $B$, respectively.
The proof of the following result is given in Section \ref{sec:appendix}.

\begin{lemma}
\label{lem:varG=varH}
In the notation above, the expectations and covariance matrices of the estimators
$\hat\theta_{G}=\int G(dt)Y(t)$ and $\hat\theta_{H}=\int H(dt)Y(t)$ coincide.
\end{lemma}

{ By Lemma  \ref{lem:varG=varH} we can restrict the search of  linear unbiased estimators to  estimators $\hat\theta_{G}$
of the form \eqref{eq:est}, where  the components $G_1, \ldots, G_q$ of the   signed matrix-measure $G=(G_0,\ldots,G_q)$
 have no continuous parts.}

\subsection{Several examples of the BLUE for non-differentiable error processes}
\label{sec:no-deriv}

For the sake of completeness we first consider the  case when the errors in model \eqref{eq:model}   follow a Markov process; this includes the case 
of
continuous autoregressive  errors of order 1.
In presenting these results we follow \cite{N1985a} and    \cite{DetPZ2016}.

\begin{proposition}
\label{prop:K=uv}
Consider the regression model \eqref{eq:model} with
  covariance kernel $K(t,s)=u(t)v(s)$ for $t\le s$ and $K(t,s)=v(t)u(s)$ for $t>s $,
where $u(\cdot)$ and $v(\cdot)$
are positive functions such that $q(t)=u(t)/v(t)$ is monotonically increasing.
Define the signed vector-measure $\zeta(dt)=z_A\delta_A(dt)+z_B\delta_B(dt)+z(t)dt$ with
\bea
 z_{A}&=&\frac{  1 }{ v^2(A) q^\prime (A)}  \Big[  \frac{f(A)u^\prime(A)}{u(A)} - f^\prime(A)  \Big]\, ,\;~~\\
 z(t)&=&-\frac{1}{ v( t )} \Big[ \frac{h^{\prime }(t)  }{q^{\prime } (t)}  \Big]^\prime\,,~~
 z_{B}=\frac{   h^\prime(B)}{ v(B) q^\prime (B)},
\eea
where  the vector-function $h(\cdot)$ is defined by $h(t)=f(t)/v(t)$.
Assume that the matrix $C=\int_\T f(t)\zeta^T(dt)$
is non-degenerate.
Then the estimate $\hat\theta_G$ with $G(dt)=C^{-1}\zeta(dt)$
is a BLUE with  covariance matrix $C^{-1}$.
\end{proposition}

In the following statement we provide an explicit expression for the BLUE in  a special case
of the covariance kernel $K(t,s)$ such that  $K(t,s)\neq u(t)v(s)$. This statement provides the first
example where  an explicit form of the BLUE and its covariance matrix can be obtained for a non-Markovian error process.
The proof is given in Section \ref{sec:appendix}.

\begin{proposition}
\label{prop:CONDIT-MARKOV}
Consider the regression model \eqref{eq:model} on the interval ${\cal T}=[A,B]$ with errors having
the covariance function $K(t,s)=1+\lambda_1 t-\lambda_2 s,$ where $t\le s$, $\lambda_1\ge\lambda_2$ and $\lambda_2(B-A)\le 1$.
Define the signed vector-measure $\zeta(dt)=z_A\delta_A(dt)+z_B\delta_B(dt)+z(t)dt$
by
\bea
 z(t)=-\frac{f^{(2)}(t)}{\lambda_1+\lambda_2},
 &&z_{A}=\Big(-f^{(1)}(A)+\frac{\lambda_1^2f(A)+\lambda_1\lambda_2f(B)}{\lambda_1+\lambda_2+\lambda_1^2A-\lambda_2^2B)}\Big)/(\lambda_1+\lambda_2),\\
 &&z_{B}=\Big(f^{(1)}(B)+\frac{\lambda_1\lambda_2f(A)+\lambda_2^2f(B)}{\lambda_1+\lambda_2+\lambda_1^2A-\lambda_2^2B)}\Big)/(\lambda_1+\lambda_2)
\eea
and suppose that the matrix
${C}\!=\!\int_\T f(t)\zeta^T(dt)$
is non-degenerate.
Then the estimator $\hat\theta_G$ with $G(dt)={C}^{-1}\zeta(dt)$
is a BLUE with  covariance matrix  ${C}^{-1}$.
\end{proposition}

If $\lambda_1=\lambda_2$ and $[A,B]=[0,1]$  in Proposition~\ref{prop:CONDIT-MARKOV}
then we obtain the following   case when the  kernel is
\be
\label{eq:triag_kernel}
K(t,s)=\max(1-\lambda|t-s|,0)\, .
\ee
Optimal designs for this covariance kernel (with $\lambda=1$) have been considered in
[Sect. 6.5 in \cite{N1985a}], \cite{muller2003measures} and \cite{fedorov2007optimum}.

\begin{example} {\rm
\label{prop:TRIANG}
Consider the regression model \eqref{eq:model} on the interval $\T=[0,1]$ with errors having
the covariance kernel \eqref{eq:triag_kernel} with $\lambda\le 1$.
Define the signed vector-measure
 $$\zeta(dt)=[-f^{(1)}(0)/(2\lambda)+f_\lambda]\delta_0(dt)+[f^{(1)}(1)/(2\lambda)+f_\lambda]\delta_1(dt)-[f^{(2)}(t)/(2\lambda)]dt\, ,$$
where $f_\lambda=  (f(0)+f(1))/(4-2\lambda)$. Assume that the matrix
$
C =\int_\T f(t)\zeta^T(dt)
$  is non-degenerate.
Then the estimator $\hat\theta_G$ with $G(dt)=C^{-1}\zeta(dt)$ is a BLUE; the  covariance matrix of this estimator is given by $C^{-1}$.
}
\end{example}

Consider now the case
when the regression functions are linear combinations of eigenfunctions from the Mercer's theorem.
Note that a similar approach was used in \cite{DetPZ2012} for the construction of optimal designs for the signed least squares estimators.
Let $\T=[A,B]$ and $\nu$ be  a  measure on the corresponding Borel field with positive density.
 Consider the integral operator
\be \label{opdef}
T_K(h)(\cdot) = \int_{A}^B K(t,\cdot )h(t) \nu (d t)
\ee
on $L_2(\nu,[A,B])$, which  defines a symmetric,
compact self-adjoint operator. 
In this case  Mercer's Theorem [see e.g. \cite{kanwal1997}]
shows that there exist a countable number of orthonormal
(with respect to $\nu (d t)$) eigenfunctions $\phi_1, \phi_2, \ldots $ with positive eigenvalues
$\lambda_1,\lambda_2,\ldots $ of the integral operator $T_K$.
The next statement  follows directly from Proposition~\ref{th:GenGMq}.

\begin{proposition}
\label{th:KL}{
Let $\phi_1, \phi_2, \ldots $ be the eigenfunctions of the integral operator  \eqref{opdef}
and
 $f(t)=\sum_{\ell=1}^\infty q_\ell\phi_{\ell}(t)$ with some sequence $\{q_\ell\}_{\ell\in \mathbb{N}}$ in $\mathbb{R}^m$
such that $f_1(t),\ldots,$ $f_m(x)$ are linearly independent.
Then the estimator $\hat\theta_G$ with $G(dt)=C^{-1}
 \sum_{\ell=1}^\infty {\lambda_\ell^{-1}}{q_\ell} \phi_{\ell}(t) \nu(dt)$ and
$C=\sum_{\ell=1}^\infty {\lambda_\ell^{-1}}{q_\ell}{q_\ell^T} $
is a BLUE
with  covariance matrix  $C^{-1}$.}
\end{proposition}

Proposition \ref{th:KL} provides a way of constructing the covariance kernels
for which the measure defining the BLUE does not have any atoms.
An example of such kernels is the following.

\begin{example} {\rm
Consider the regression model \eqref{eq:model} with $m=1$, $f(t)\equiv 1$, $t\in\T=[-1,1]$,
and the covariance kernel
$
 K(t,s)=1+\kappa p_{\alpha,\beta}(t)p_{\alpha,\beta}(s),
$
where $\kappa>0,\alpha,\beta>-1$ are some constants and
$p_{\alpha,\beta}(t)=\frac{\alpha-\beta}2+(1+\frac{\alpha+\beta}2)t$
is the Jacobi polynomial of degree $1$.
Then the estimator $\hat\theta_G$ with $G(dt)={\rm const} \cdot (1-t)^\alpha(1+t)^\beta dt$ is a BLUE.
}
\end{example}

\section{BLUE for   processes  with trajectories  in $C^1[A,B]$}
\label{sec:BLUE-C1}
\def\theequation{3.\arabic{equation}}
\setcounter{equation}{0}


In this section, we assume that the error process is exactly once  continuously differentiable (in the mean-square sense).

\subsection{A general statement}
Consider the regression model \eqref{eq:model}
and a linear estimator in the form
\be \label{estc1}
 \hat\theta_{G_0,G_1}=\int_\T y(t)G_0(dt)+\int_\T y^{(1)}(t)G_1(dt),
\ee
where $G_0(dt)$ and $G_1(dt)$ are signed vector-measures.
%
The following corollary is a specialization  of Proposition~\ref{th:GenGMq} when $q=1$.

\begin{corollary}
\label{cor:GGM-1}
Consider the regression model \eqref{eq:model} with the covariance kernel $K(t,s)$ and
such that  $y^{(1)}(t)$ exists in the mean-square sense for all $t  \in [A,B]$.
Suppose that  $y(t)$ and $y^{(1)}(t)$ can be observed at all $t\in \T$.
Assume that there exist vector-measures $\zeta_0$ and $\zeta_1$ such that
the equality
\bea
 \int_\T K(t,s)\zeta_0(dt)+\int_\T  K^{(1)}(t,s)\zeta_1(dt)=f(s),
\eea
is fulfilled for all $s\in\T$, and such that
  the matrix
  $$C=\int_\T f(t)\zeta_0^T(dt)+\int_\T f^{(1)}(t)\zeta_1^T(dt)$$
  is non-degenerate.
Then the estimator $\hat\theta_{G_0,G_1}$ defined in \eqref{estc1} with $G_i=C^{-1}\zeta_i$ ($i=0,1$) is a BLUE
with  covariance matrix ${C}^{-1}$.
\end{corollary}

 The next  theorem  provides  sufficient conditions for vector-measures of some particular form
 to define a BLUE by \eqref{estc1} for the case $\T=[A,B]$.
 This theorem, which is proved in Section \ref{sec:appendix}, will be useful for several choices of the covariance kernel below.
Define the vector-function
\bea
z(t)&=&(\tau_0f(t)-\tau_2f^{(2)}(t)+ f^{(4)}(t))/s_3,
\eea
and vectors
\bea
z_A&=&\big(f^{(3)}(A)- \gamma_{1,A} f^{(1)}(A)  + \gamma_{0,A} f(A)\big)/s_3,\\
z_B&=&\big(-f^{(3)}(B)+ \gamma_{1,B} f^{(1)}(B)  + \gamma_{0,B} f(B)\big)/s_3,\\
z_{1,A}&=&\big(-f^{(2)}(A)+\beta_{1,A}f^{(1)}(A)- \beta_{0,A} f(A)\big)/s_3,\\
z_{1,B}&=&\big(f^{(2)}(B)+\beta_{1,B}f^{(1)}(B)+ \beta_{0,B} f(B)\big)/s_3,
\eea
where $\tau_0,\tau_2,\gamma_{0,A},\gamma_{1,A},\beta_{0,A},\beta_{1,A},\gamma_{0,B},\gamma_{1,B},\beta_{0,B},\beta_{1,B},s_3$
are some constants and $s_3=K^{(3)}(s-,s)-K^{(3)}(s+,s).$
Define the functions
\be
 \label{eq:gam-beta-1} ~~~~~~~~
\begin{aligned}
 J_1(s)=&-\gamma_{1,A}K(A,s)+\beta_{1,A}K^{(1)}(A,s)+\tau_2 K(A,s)- K^{(2)}(A,s),\\
 J_2(s)=&~~\gamma_{0,A}K(A,s)-\beta_{0,A}K^{(1)}(A,s)-\tau_2K^{(1)}(A,s)+ K^{(3)}(A,s),\\
 J_3(s)=&-\gamma_{1,B}K(B,s)+\beta_{1,B}K^{(1)}(B,s)-\tau_2 K(B,s)+ K^{(2)}(B,s),\\
 J_4(s)=&~~\gamma_{0,B}K(B,s)-\beta_{0,B}K^{(1)}(B,s)+\tau_2K^{(1)}(B,s)- K^{(3)}(B,s).
\end{aligned}
\ee

\begin{theorem}
\label{th:BLUE-2}
Consider the regression model \eqref{eq:model} on the interval ${\cal T}=[A,B]$ with errors having the covariance kernel $K(t,s)$.
Suppose that the vector of regression functions $f$ is four times differentiable and $K(t,s)$ is also four times differentiable for $t\neq s$ such that
\bea
 K^{(i)}(s-,s)&-&K^{(i)}(s+,s)=0, ~~i=0,1,2,\\
 K^{(3)}(s-,s)&-&K^{(3)}(s+,s)\neq0.
\eea
With the notation of the previous paragraph define the vector-measures
\begin{eqnarray*}
\zeta_0(dt) &=& z_A\delta_A(dt)+z_B\delta_B(dt)+z(t)dt ,\\
\zeta_1(dt) &=& z_{1,A}\delta_A(dt)+z_{1,B}\delta_B(dt).
\end{eqnarray*}
Assume that there exist constants $\tau_0,\tau_2,\gamma_{0,A},$ $\gamma_{1,A},\beta_{0,A},$ $\beta_{1,A},$ $\gamma_{0,B},\gamma_{1,B},\beta_{0,B},\beta_{1,B}$
such that
(i)  the identity
\be
 \tau_0 K(t,s)-\tau_2 K^{(2)}(t,s)+ K^{(4)}(t,s)\equiv 0
 \label{eq:tau4}
\ee
holds for all $t,s\in[A,B]$,
(ii) the identity $J_1(s)+J_2(s)+J_3(s)+J_4(s)\equiv 0$
holds for all $s\in[A,B]$, and
(iii) the matrix
${C}=\int_\T f(t)\zeta_0^T(dt)+\int_\T f^{(1)}(t)\zeta_1^T(dt)$ is non-degenerate. Then the estimator $\hat\theta_{G_0,G_1}$ defined in \eqref{estc1} with $G_i(dt)={C}^{-1}\zeta_i(dt)$ $(i=0,1)$ is a BLUE
with  covariance matrix  ${C}^{-1}$.
\end{theorem}

\subsection{Two examples for integrated error processes}
In this section we illustrate the application of our results calculating the  BLUE when errors follow
an integrated Brownian motion and an integrated  process with triangular-shape kernel.
All results of this section can  be verified  by a direct application of Theorem~\ref{th:BLUE-2}.
We first consider the case of Brownian motion, where the integrated covariance kernel is given by
\begin{eqnarray} \nonumber
 K(t,s) &=& \int_a^t\int_a^s \min(t',s')dt'ds' \\
&=& \frac{\max(t,s)(\min(t,s)^2-a^2)}{2}-\frac{a^2(\min(t,s)-a)}{2}-\frac{\min(t,s)^3-a^3}{6} \label{intBB}
 \end{eqnarray}
and $0 \leq a \leq A$.

\begin{proposition}
\label{prop:IBMa}
Consider the regression model \eqref{eq:model} with integrated covariance kernel given by \eqref{intBB}
and suppose that $f$ is four times differentiable on the interval $[A,B]$.
Define
the signed  vector-measures
\begin{eqnarray*}
\zeta_0(dt) &=& z_A\delta_A(dt)+z_B\delta_B(dt)+z(t)dt, \\
\zeta_1(dt) &=& z_{1,A}\delta_A(dt)+z_{1,B}\delta_B(dt),
\end{eqnarray*}
where $z(t)=f^{(4)}(t)$,
\bea
 z_A&=&f^{(3)}(A)-\frac{6(A+a)}{(A+3a)(A-a)^2}f^{(1)}(A)+\frac{12A}{(A+3a)(A-a)^3}f(A),\\
 z_{1,A}&=&-f^{(2)}(A)+\frac{4(A+2a)}{(A+3a)(A-a)}f^{(1)}(A)-\frac{6(A+a)}{(A+3a)(A-a)^2}f(A),\\
 z_B&=&-f^{(3)}(B),~~~
 z_{1,B}=f^{(2)}(B).
\eea
Assume that the matrix
${C}=\int_A^B f(t)\zeta_0^T(dt)+\int_\T f^{(1)}(t)\zeta_1^T(dt)$ is non-degenerate.
Then the estimator $\hat\theta_{G_0,G_1}$ defined in \eqref{estc1} with $G_i(dt)={C}^{-1}\zeta_i(dt)$ is a BLUE
with covariance matrix ${C}^{-1}$.
\end{proposition}

The next example is a particular case of Proposition \ref{prop:IBMa} when $a=0$.

\begin{example} {\rm
\label{ex:IBM}
Consider the regression model \eqref{eq:model} on $\T=[A,B]$ with integrated covariance kernel
\be
\label{eq:int_cov_kern}
 K(t,s)={\min(t,s)^2}(3\max(t,s)-\min(t,s))/{6} \, .
\ee
Suppose that $f$ is differentiable four times.
Define
the vector-measures $\zeta_0(dt)=z_A\delta_A(dt)+z_B\delta_B(dt)+z(t)dt$ and
$\zeta_1(dt)=z_{1,A}\delta_A(dt)+z_{1,B}\delta_B(dt)$, where $z(t)=f^{(4)}(t)$,
\bea
 z_A&=&f^{(3)}(A)-\frac{6}{A^2}f^{(1)}(A)+\frac{12}{A^3}f(A),\\
 z_{1,A}&=&-f^{(2)}(A)+\frac{4}{A}f^{(1)}(A)-\frac{6}{A^2}f(A),\\
  z_B&=&-f^{(3)}(B),~~~ z_{1,B}=f^{(2)}(B).
\eea
Assume that the matrix
 ${C}=\int_A^B f(t)\zeta_0^T(dt)+\int_A^B f^{(1)}(t)\zeta_1^T(dt)$ is non-degenerate.
Then the estimator $\hat\theta_{G_0,G_1}$ defined in \eqref{estc1} with $G_i(dt)={C}^{-1}\zeta_i(dt)$ is a BLUE
with covariance matrix  ${C}^{-1}$.
}
\end{example}

Consider now the integrated triangular-shape kernel
\begin{eqnarray} \nonumber
 K(t,s) &=& \int_0^t\int_0^s\max\{0,1-\lambda|t'-s'|\}dt'ds' \\
  &=& ts-\lambda\min(t,s)\Big(3\max(t,s)^2-3ts+2\min(t,s)^2\Big)/6.
 \label{eq:integ-trK}
\end{eqnarray}

\begin{proposition}
Consider the regression model \eqref{eq:model} on $\T=[A,B]$ with integrated covariance kernel \eqref{eq:integ-trK},
where $\lambda(B-A)<1$.
Suppose that $f$ is four times differentiable.
Define
the signed vector-measures
\begin{eqnarray*}
\zeta_0(dt) &=&z_A\delta_A(dt)+z_B\delta_B(dt)+z(t)dt ,  \\
\zeta_1(dt) &=& z_{1,A}\delta_A(dt)+z_{1,B}\delta_B(dt),
\end{eqnarray*}
 where
$z(t)=f^{(4)}(t)/(2\lambda)$
and
\bea
 z_A&=&\Big[f^{(3)}(A)-\frac{6\kappa_{2}}{A^2\kappa_{4}}f^{(1)}(A)+\frac{6\lambda}{A\kappa_{4}}f^{(1)}(B)+\frac{12\kappa_{1}}{A^3\kappa_{4}}f(A)\Big]/(2\lambda),\\
 z_{1,A}&=&\Big[-f^{(2)}(A)+\frac{4\kappa_{3}}{A\kappa_{4}}f^{(1)}(A)-\frac{2\lambda}{\kappa_{4}}f^{(1)}(B)-\frac{6\kappa_{2}}{A^2\kappa_{4}}f(A)\Big]/(2\lambda),\\
 z_{1,B}&=&\Big[f^{(2)}(B)-\frac{2\lambda}{\kappa_{4}}f^{(1)}(A)+\frac{4\lambda}{\kappa_{4}}f^{(1)}(B)+\frac{6\lambda}{A\kappa_{4}}f(A)\Big]/(2\lambda),\\
 z_B&=&-f^{(3)}(B)/(2\lambda),~~\kappa_{j}=A\lambda-j B\lambda+2j.
\eea
Assume that the matrix
$C=\int_A^B f(t)\zeta_0^T(dt)+\int_A^B f^{(1)}(t)\zeta_1^T(dt)$ is non-degenerate.
Then the estimator $\hat\theta_{G_0,G_1}$ defined in \eqref{estc1} with  $G_i(dt)={C}^{-1}\zeta_i(dt)$ is a BLUE
with  covariance matrix  $C^{-1}$.
\end{proposition}

\subsection{Explicit form of the BLUE for the integrated processes}
\label{sec:SYkernels}

We conclude this section establishing  a direct link between the BLUE for models with
non-differentiable error processes and the BLUE for  regression models with an
 integrated kernel  \eqref{SYkernel}. Note that this extends the class of kernels considered in \cite{SY1970} in a nontrivial way.

Consider the regression model
\eqref{eq:model} with  a non-differentiable error process
with covariance kernel $K(t,s) $ and   BLUE
$$
\hat \theta_{G_0} = \int_\T y(t) G_0(dt).
$$
From Proposition~\ref{th:GenGMq} we have for the vector-measure $\zeta_0(dt)$ satisfying \eqref{eq:suff-cond-q} and defining the BLUE
\be
 \label{26neu}
 \int_{A}^B K(t,s)\zeta_0(dt)=  f(s)
\ee
and
 $ \mathrm{Var}(\hat \theta_{G_0})=C^{-1}=\big(\int_\T f(t)\zeta_0^T(dt)\big)^{-1}$.
The unbiasedness condition for the measure $G_0(dt)=C^{-1}\zeta_0(dt)$ is
\bea
\int_\T f(t) G^T_0(dt)=I_m.
\eea

Define the integrated process as follows:
$$
\widetilde{y}(t)=\int_a^t y(u) du,\;\;\
\widetilde{f}(t)=\int_a^t f(u) du,\;\;\
\widetilde{\varepsilon}(t)=\int_a^t \varepsilon(u) du\;\;\
$$
with some $a\le A$ (meaning that the regression vector-function and the error process are defined on $[a,B]$ but observed on $[A,B]$)
so that
$$
\widetilde{f}^{(1)}(t)= f(t),\;\;\
\widetilde{y}^{(1)}(t)= y(t),\;\;\
\widetilde{\varepsilon}^{(1)}(t)= \varepsilon(t) \; .
$$
Consider the regression model
\be
\label{model2}
\tilde{y}(t)=\theta^T \tilde{f}(t)+ \tilde{\varepsilon}(t),
\ee
which has the integrated covariance kernel
\be
\label{SYkernel}
R(t,s)=\int_a^t \int_a^{s} K(u,v) du dv.
\ee
The  proof of the following result is given in Section \ref{sec:appendix}.

\begin{theorem}
\label{th:integr}
Let the vector-measure $\zeta_0$ satisfy the equality \eqref{26neu} and
define the BLUE $\hat\theta_{G_0}$  with $G_0(dt)=C^{-1}\zeta_0(dt)$ in the regression model \eqref{eq:model}
with covariance kernel $K(\cdot,\cdot)$.
Let the measures $\eta_0,\eta_1$ satisfy the equality
\be
\label{thm32cond}
 \int_\T  R(t,s)\eta_0(dt)+ \int_\T  R^{(1)}(t,s)\eta_1(dt)= 1
 \ee
for all $s \in \T$.
{Define the vector-measures
$\tilde\zeta_0=- c\eta_0$ and $\tilde\zeta_1=- c\eta_1+\zeta_0$,}
where the vector $c$ is given by
$c=\int_{a}^A[\int_A^B K(t,s)\zeta_0(dt)-  f(s)]ds$.
Then
the estimator $\hat\theta_{\tilde G_0,\tilde G_1}$ defined in \eqref{estc1} with $\tilde G_i(dt)=\tilde C^{-1}\tilde \zeta_i(dt)$ $(i=1,2)$,
where $\tilde C=\int \tilde f(t)\tilde \zeta_0^T(dt)+\int \tilde f^{(1)}(t)\tilde \zeta_1^T(dt)$, is a BLUE in the regression model \eqref{model2}
with integrated covariance kernel~\eqref{SYkernel}.
\end{theorem}

{ Repeated application of Theorem~\ref{th:integr} extends
the results to the case  of several times integrated processes.}

If   $a=A$ in   \eqref{SYkernel}  we have  $c=0$ in Theorem \ref{th:integr} and in this case,
the statement of Theorem~\ref{th:integr} can be  proved  easily.
Moreover,  in this case  the class of kernels defined by  \eqref{SYkernel}  is exactly the class of kernels
considered in equation (1.5) and (1.6)  of  \cite{SY1970} for once differentiable processes ($k=1$ in their notation).
We emphasize that the class of kernels
considered here is much richer than the class of kernels considered in \cite{SY1970}.

\subsection{BLUE for AR(2) errors}
\label{sec:AR2}


Consider the continuous-time regression model \eqref{eq:model},
which can be observed at all $t\in[A,B]$, where the error process is   a continuous autoregressive (CAR) process of order $2$.
Formally, a CAR$(2)$ process is defined as  a solution of the linear stochastic differential equation of the form
\be \label{car}
d \varepsilon^{(1)}(t)=\tilde a_1\varepsilon^{(1)}(t)+\tilde a_2\varepsilon(t)+\sigma^2_0dW(t),
\ee
 where $\mathrm{Var}(\varepsilon(t))=\sigma^2$ and $W(t)$ is a standard Wiener process,
[see \cite{BDY2007}].
Note that the process $\{\varepsilon(t)| t \in [A,B]\}$ defined by \eqref{car} has a continuous derivative  and, consequently,
the process $ \{ y(t) = \theta^T f(t) + \varepsilon (t) |~ t  \in [A,B] \}$,
is a continuously differentiable process with drift  on the interval $ [A,B]$. In this section we derive the explicit form for the continuous BLUE using 
Theorem~\ref{th:BLUE-2}. An alternative approach would be to use the coefficients of the equation \eqref{car} as indicated in \cite{parzen1961approach}.

There are in fact three different  forms of the  autocorrelation function $\rho(t)=K(0,t)$
of  CAR(2) processes [see e.g. formulas (14)--(16) in \cite{he1989embedding}], which are given by
\be
 \rho_{1}(t)=\frac{\lambda_2}{\lambda_2-\lambda_1} e^{-\lambda_1|t|}-\frac{\lambda_1}{\lambda_2-\lambda_1} e^{-\lambda_2|t|}\, ,
\label{eq:K-eeC}
\ee
 where $\lambda_1\neq\lambda_2$, $\lambda_1>0$, $\lambda_2>0$, by 
\be
\rho_{2}(t)=e^{-\lambda |t|}\Big\{\cos( \omega |t|)+ \frac{\lambda}{\omega} \sin(\omega |t|)\Big\}\,,
\label{eq:K-ecosC}
\ee
where $\lambda>0$, $\omega>0$, and by
\be
\rho_{3}(t)=e^{-\lambda |t|}(1+ \lambda |t|)\, ,
\label{eq:K-elinC}
\ee
where $\lambda>0$. Note that the
kernel \eqref{eq:K-elinC} is widely known as  Mat\'{e}rn kernel with parameter 3/2,
which has numerous applications in spatial statistics  [see \cite{rasmussen2006gaussian}] and computer experiments  [see \cite{pronzato2012design}].
In the following results, which are proved in Section \ref{sec:last},  we specify the BLUE for the CAR($2$) model.


\begin{proposition}
\label{prop:CAR1-expexp}
Consider the regression model \eqref{eq:model} with CAR(2) errors, where   the covariance kernel $K(t,s)=\rho(t-s)$ has the form \eqref{eq:K-eeC}.
Suppose that $f$ is a vector  of linearly independent, four times differentiable  functions on the interval $[A,B]$.
Then the conditions of Theorem \ref{th:BLUE-2} are satisfied for
$s_3=2\lambda_1\lambda_2(\lambda_1+\lambda_2)$,
$\tau_0=\lambda_1^2\lambda_2^2$, $\tau_2=\lambda_1^2+\lambda_2^2$,
$\beta_{j,A}=\beta_{j,B}=\beta_j$ and
$\gamma_{j,A}=\gamma_{j,B}=\gamma_j$ for $j=0,1$,
where
$\beta_1=\lambda_1+\lambda_2$, $\gamma_1=\lambda_1^2+\lambda_1\lambda_2+\lambda_2^2$,
$\beta_0=\lambda_1\lambda_2$ and $\gamma_0=\lambda_1\lambda_2(\lambda_1+\lambda_2)$.
\end{proposition}

\begin{proposition}
\label{prop:CAR1-expcos}
Consider the regression model \eqref{eq:model} with CAR(2) errors, where   the covariance kernel $K(t,s)=\rho(t-s)$ has the form \eqref{eq:K-ecosC}.
Suppose that $f$ is a vector  of linearly independent, four times differentiable  functions.
Then the conditions of Theorem \ref{th:BLUE-2} hold for
$s_3=4\lambda(\lambda^2+\omega^2)$,
$\tau_0= (\lambda^2+\omega^2)^2$, $\tau_2=2(\lambda^2-\omega^2)$,
$\beta_{j,A}=\beta_{j,B}=\beta_j$ and
$\gamma_{j,A}=\gamma_{j,B}=\gamma_j$ for $j=0,1$,
where
$\beta_1=2\lambda$, $\gamma_1=\gamma_1= 3\lambda^2-\omega^2$,
$\beta_0=\lambda^2+\omega^2$ and $\gamma_0=2\lambda(\lambda^2+\omega^2)$.
\end{proposition}

The BLUE for the covariance kernel in the form \eqref{eq:K-elinC}
is obtained from either Proposition \ref{prop:CAR1-expexp} with $\lambda_1=\lambda_2=\lambda$
or Proposition \ref{prop:CAR1-expcos} with $\omega=0$.

\begin{remark} {\rm
In the online supplement \cite{supplement}
 we consider the regression model \eqref{eq:modeldisc} with a discrete AR($2$)
error process. Although  the discretised CAR($2$) process follows  an ARMA($2,1$) model rather than
an AR($2$) [see \cite{he1989embedding}] we will be able to
 establish the  connection between the BLUE in the discrete and   continuous-time models and hence derive  the limiting form
 of the  discrete BLUE and  its covariance matrix. }
\end{remark}

\section{Models with more than once differentiable error processes}
\label{sec:ARp+IBM}
\def\theequation{4.\arabic{equation}}
\setcounter{equation}{0}

If $\T=[A,B]$ and $q>1$ then solving the Wiener-Hopf type equation \eqref{eq:suff-cond-q}
numerically is virtually impossible  in view of the fact that the problem is severely ill-posed.
Derivation of explicit forms of the BLUE for smooth kernels with $q>1$ is hence extremely important.
We did not find any general results on the form of the BLUE in such cases.
In particular, the well-known paper \cite{SY1970} dealing with these kernels does not contain any specific examples.
In Theorem \ref{th:integr} we have already established a general result
that can be used for deriving explicit forms for the BLUE for $q>1$ times integrated kernels,
which  can be used repeatedly for this purpose.
We can also formulate  a result similar to Theorem~\ref{th:BLUE-2}.
However, already for $q=2$, even a formulation of such theorem  would take a couple of pages and hence its usefulness would be very doubtful.

In this section, we indicate how the general methodologies developed in the previous sections
can be extended to    error processes with $q>1$  by two examples: twice integrated Brownian motion and CAR($p$) error models
with $p\geq 3$, but other cases can be treated very similarly.

\subsection{Twice integrated Brownian motion}
\label{sec:double}

\begin{proposition}
Consider the regression model \eqref{eq:model} where the error process is the twice integrated Brownian motion with the covariance kernel
\bea
 K(t,s)=t^5/5!-st^4/4!+s^2t^3/12,~t<s.
\eea
Suppose that $f$ is $6$ times differentiable  and define
the vector-measures
\bea
\zeta_0(dt) &=& z_A\delta_A(dt)+z_B\delta_B(dt)+z(t)dt, \\
\zeta_1(dt)&=& z_{1,A}\delta_A(dt)+z_{1,B}\delta_B(dt), \\
\zeta_2(dt) &=&z_{2,A}\delta_A(dt)+z_{2,B}\delta_B(dt),
\eea
 where $ z(t)=f^{(6)}(t)$,
\bea
 z_A&=&(A^5 f^{(5)}(A)-60A^2f^{(2)}(A)+360Af^{(1)}(A)-720f(A))/A^5,\\
 z_{1,A}&=&-(A^4 f^{(4)}(A)-36A^2f^{(2)}(A)+192Af^{(1)}(A)-360f(A))/A^4,\\
 z_{2,A}&=&(A^3f^{(3)}(A)-9A^2f^{(2)}(A)+36Af^{(1)}(A)-60f(A))/A^3,\\
 z_B&=& -f^{(5)}(B),~~ z_{1,B}=f^{(4)}(B),~~ z_{2,B}=-f^{(3)}(B).
\eea
Then the estimator $\hat\theta_{G_0,G_1,G_2}$ defined by \eqref{eq:est} (for $q=2$) with $G_i(dt)={C}^{-1}\zeta_i(dt)$ $(i=0,1,2)$,
$$
{C}=\int_\T f(t)\zeta_0^T(dt)+\int_\T f^{(1)}(t)\zeta_1^T(dt)+\int_\T f^{(2)}(t)\zeta_2^T(dt),$$  is the BLUE
with  covariance matrix  ${C}^{-1}$.
\end{proposition}

\subsection{CAR(p) models with p$ \geq 3$}

\label{sec:carp}

Consider the regression model \eqref{eq:model},
which can be observed at all $t\in[A,B]$ and the error process has the continuous autoregressive (CAR) structure of order $p$.
Formally, a CAR$(p)$ process is a solution of the linear stochastic differential equation of the form
$$
d \varepsilon^{(p-1)}(t)=\tilde a_1\varepsilon^{(p-1)}(t)+\ldots+\tilde a_p\varepsilon(t)+\sigma^2_0dW(t),
$$
 where $\mathrm{Var}(\varepsilon(t))=\sigma^2$ and $W$ is a standard Wiener process,
[see \cite{BDY2007}].
Note that the process $\varepsilon$ has continuous derivatives $\varepsilon^{(1)}(t),\ldots,$ $\varepsilon^{(p-1)}(t)$ at the point $t$ and, consequently,
the process $ \{ y(t) = \theta^T f(t) + \varepsilon (t) |~ t  \in [A,B] \}$
  is continuously differentiable $p-1$ times on the interval $ [A,B]$ with drift $\theta^T f(t)$.
Define the vector-functions
\bea
z(t)&=&(\tau_0f(t)+\tau_2f^{(2)}(t)+\ldots+ f^{(2p)}(t))/s_{2p-1},
\eea
and vectors
\bea
z_{j,A}&=&\sum\nolimits_{l=0}^{2p-j-1} \gamma_{l,j,A} f^{(j)}(A)/s_{2p-1},\\
z_{j,B}&=&\sum\nolimits_{l=0}^{2p-j-1} \gamma_{l,j,B} f^{(j)}(B)/s_{2p-1}
\eea
for $j=0,1,\ldots,p-1$, where $s_{2p-1}=K^{(2p-1)}(s-,s)-K^{(2p-1)}(s+,s).$

\begin{proposition}
\label{prop:q>1}
Consider the regression model \eqref{eq:model} with CAR$(p)$ errors.
Define the vector-measures
\bea
\zeta_0(dt)&=&z_{0,A}\delta_A(dt)+z_{0,B}\delta_B(dt)+z(t)dt,\\
\zeta_j(dt)&=&z_{j,A}\delta_A(dt)+z_{j,B}\delta_B(dt),~j=1,\ldots,p-1,
\eea
for $j=1,\ldots,p-1$.
Then there exist constants $\tau_0,\tau_2\ldots,\tau_{2(p-1)}$
and $\gamma_{l,j,A},\gamma_{l,j,B}$, such that
the estimator $\hat\theta_{G_0,G_1,\ldots,G_{p-1}}$ defined by \eqref{eq:est} (for $q=p-1$) with $G_j(dt)={C}^{-1}\zeta_j(dt)$ $(i=0,1,\ldots,p-1)$,
$${C}=\int_\T f(t)\zeta_0^T(dt)+\sum_{j=1}^{p-1}\int_\T f^{(j)}(t)\zeta_j^T(dt),$$   is a BLUE
with covariance matrix   $C^{-1}$.
\end{proposition}

Let us consider the construction of a  BLUE for model \eqref{eq:model} with a CAR($3$) error process in more detail.
One of several possible forms for the covariance function for the CAR($3$) process is given by
\be
 \rho(t)=
  c_1 e^{-\lambda_1|t|}+ c_2 e^{-\lambda_2|t|}+ c_3 e^{-\lambda_3|t|}\, ,
\label{eq:K3eeC}
\ee
where $\lambda_1, \lambda_2, \lambda_3$ are the roots of the autoregressive polynomial $\tilde a(z)=z^3+\tilde a_1z^2+\tilde a_2z+\tilde a_3$,
$$
 c_j=\frac{k_j}{k_1+k_2+k_3},~~k_j=\frac{1}{\tilde a'(\lambda_j)\tilde a(-\lambda_j)},
$$
$\lambda_i\neq\lambda_j$, $\lambda_i>0$, $i,j=1,\ldots,3$, see \cite{brockwell2001continuous}.
Specifically, we have
\bea
 c_1=\frac{\lambda_2\lambda_3(\lambda_2+\lambda_3)}{(\lambda_1-\lambda_2)(\lambda_1-\lambda_3)(\lambda_1+\lambda_2+\lambda_3)},\\
 c_2=\frac{\lambda_1\lambda_3(\lambda_1+\lambda_3)}{(\lambda_2-\lambda_1)(\lambda_2-\lambda_3)(\lambda_1+\lambda_2+\lambda_3)},\\
 c_3=\frac{\lambda_1\lambda_2(\lambda_1+\lambda_2)}{(\lambda_3-\lambda_1)(\lambda_3-\lambda_2)(\lambda_1+\lambda_2+\lambda_3)}.
\eea

In this case, a BLUE is given  in Proposition~\ref{prop:q>1} with the following parameters:
\bea
 \tau_0=-\lambda_1^2\lambda_2^2\lambda_3^2,~
 \tau_2=\lambda_1^2\lambda_2^2+\lambda_1^2\lambda_3^2+\lambda_2^2\lambda_3^2,~
 \tau_4=-\lambda_1^2-\lambda_2^2-\lambda_3^2,
\eea
\bea
 s_5=\frac{2\lambda_1\lambda_2\lambda_3(\lambda_1+\lambda_2)(\lambda_1+\lambda_3)(\lambda_2+\lambda_3)}{\lambda_1+\lambda_2+\lambda_3}
 =2\frac{\prod_i \lambda_i \prod_{i\neq j}(\lambda_i+\lambda_j)}{\sum_i \lambda_i},
\eea
\bea
 z_{0,A}&=&f^{(5)}(A)-\tsum_i \lambda_i^2f^{(3)}(A)-\tprod_i \lambda_if^{(2)}(A)\\
 &&+[\tsum_{i\neq j}\lambda_i^2\lambda_j^2+\tprod_i \lambda_i\tsum_i \lambda_i]f^{(1)}(A)-\tprod_i \lambda_i \tsum_{i\neq j} \lambda_i\lambda_j f(A)\\
 z_{1,A}&=&-f^{(4)}(A)+\tsum_{i, j}\lambda_i\lambda_j f^{(2)}(A)-
 \tprod_{i\neq j}(\lambda_i+\lambda_j) f^{(1)}(A)+\tprod_i \lambda_i \tsum_i \lambda_i f(A)\\
 z_{2,A}&=&f^{(3)}(A)-\tsum_i \lambda_if^{(2)}(A)+\tsum_{i\neq j} \lambda_i\lambda_j f^{(1)}(A)-\tprod_i \lambda_if(A)\\
\eea
\bea
 -z_{0,B}&=&f^{(5)}(B)-\tsum_i \lambda_i^2f^{(3)}(B)-\tprod_i \lambda_if^{(2)}(B)\\
 &&+[\tsum_{i\neq j}\lambda_i^2\lambda_j^2+\tprod_i \lambda_i\tsum_i \lambda_i]f^{(1)}(B)-\tprod_i \lambda_i \tsum_{i\neq j} \lambda_i\lambda_j f(B)\\
 -z_{1,B}&=&-f^{(4)}(B)+\tsum_{i, j}\lambda_i\lambda_j f^{(2)}(B)-
 \tprod_{i\neq j}(\lambda_i+\lambda_j) f^{(1)}(B)+\tprod_i \lambda_i \tsum_i \lambda_i f(B)\\
 -z_{2,B}&=&f^{(3)}(B)-\tsum_i \lambda_if^{(2)}(B)+\tsum_{i\neq j} \lambda_i\lambda_j f^{(1)}(B)-\tprod_i \lambda_if(B)\\
\eea

If we set $\lambda_1=\lambda_2=\lambda_3=\lambda$  then the above formulas give the explicit form of the BLUE for the
  Mat\'{e}rn kernel with parameter 5/2; that is, the kernel defined by $
\rho(t)= \left(1+\sqrt{5}t{\lambda}+{5t^2\lambda^2}/{3} \right)\exp\left(-{\sqrt{5}t\lambda}\right).
$

\section{Numerical study}
\label{sec:numer}
\def\theequation{5.\arabic{equation}}
\setcounter{equation}{0}

In this section, we describe some numerical results on comparison of  the accuracy of various estimators
for the parameters in the   regression models \eqref{eq:model} with $[A,B]=[1,2]$ and the integrated Brownian motion as   error process.
The kernel $K(t,s)$ is given in  \eqref{eq:int_cov_kern} and the explicit form of the covariance matrix of the continuous BLUE  can be found in Example~\ref{ex:IBM}. We denote this estimator by  $\hat\theta_{cont.BLUE}$. We are interested in the efficiency of various estimators for this differentiable error process.
For a given $N$ (in the tables, we use $N=3,5,10$), we consider the following four estimators that use $2N$ observations:
\begin{itemize}
  \item $\hat\theta_{BLUE } (N,N)$:  discrete BLUE based on observations  $y(t_1),\ldots,y(t_N),$ $y'(t_1),\ldots,y'(t_N)$ with $t_i=1+(i-1)/(N-1)$, $i=1,\ldots,N$.
  {This estimator uses
  $N$ observations of the original process and its derivative (at equidistant points).}
  \item $\hat\theta_{BLUE}(2N-2,2)$: discrete BLUE based on observations  $y(t_1),\ldots,$ $y(t_{2N-2}),y'(1),y'(2)$ with $t_i=1+(i-1)/(2N-3)$, $i=1,\ldots,2N-3$.
 { This estimator uses   $2N-2$ observations of the original process (at equidistant points) and  observations of its derivative at the boundary points of the design space.}
  \item $\hat\theta_{BLUE} (2N,0)$: discrete BLUE based on observations  $y(t_1),\ldots,y(t_{2N})$ with $t_i=1+(i-1)/(2N-1)$, $i=1,\ldots,2N$.
  {This estimator uses
  $2N$ observations of the original process (at equidistant points) and no observations from  its derivative.}
  \item $\hat\theta_{OLSE} (2N,0)$: ordinary least square estimator (OLSE) based on observations  $y(t_1),\ldots,y(t_{2N})$ with $t_i=1+(i-1)/(2N-1)$, $i=1,\ldots,2N$.
 { This estimator uses
  $2N$ observations of the original process (at equidistant points) and no observations from  its derivative.}
\end{itemize}
In Table \ref{tab1} -- \ref{tab3} we use the results derived in this paper to calculate the
efficiencies
 \be
 \mathrm{Eff}(\tilde\theta)={ \mathrm{Var}(\hat\theta_{cont.BLUE}) \over \mathrm{Var}(\tilde\theta)} ,
 \label{eff}
 \ee
 where $\tilde\theta$ is one of the four estimators under consideration. In particular we consider three  different scenarios  for the drift in model \eqref{eq:model}
 defined by
 \be
 \label{m1} && m=1, ~f(t)=1  \\
  \label{m2}
 && m=3, ~f(t)=(1,\sin(3\pi),\cos(3\pi))^T \\
   \label{m3}
 && m=5 , ~ f(t)=(1,t,t^2,1/t,1/t^2)^T
 \ee
\begin{table}[!hhh]
\caption{\it  The efficiency defined  by Ê\eqref{eff}   for  four
different estimators based on   $2N$ observations. The drift function is given by  \eqref{m1} }
\begin{center}
\begin{tabular}{|l|c|c|c|}
\hline
 $~~~~\tilde\theta^{\rule{0mm}{2.5mm}}$ & $N=3$ & $N=5$ & $N=10$ \\
\hline
$\hat\theta_{BLUE } (N,N) $ &1     &1     &1\\
$\hat\theta_{BLUE} (2N-2,2)$&1     &1     &1\\
$\hat\theta_{BLUE} (2N,0)$  &0.8593&0.9147&0.9570\\
$\hat\theta_{OLSE} (2N,0)$  &0.0732&0.0733&0.0734\\
\hline
\end{tabular}
\end{center}
\label{tab1}
\end{table}

\begin{table}[!hhh]
\caption{\it  \it  The efficiency defined  by Ê\eqref{eff}   for  four
different estimators based on   $2N$ observations. The drift function is given by  \eqref{m2}
}
\begin{center}
\begin{tabular}{|l|c|c|c|}
\hline
 $~~~~\tilde\theta^{\rule{0mm}{2.5mm}}$ & $N=3$ & $N=5$ & $N=10$ \\
\hline
$\hat\theta_{BLUE } (N,N) $ &0.41246&0.92907&0.99680\\
$\hat\theta_{BLUE} (2N-2,2)$&0.45573&0.98706&0.99972\\
$\hat\theta_{BLUE} (2N,0)$  &0.47796&0.77195&0.89641\\
$\hat\theta_{OLSE} (2N,0)$  &0.00113&0.00137&0.00218\\
\hline
\end{tabular}
\end{center}
\label{tab2}
\end{table}

\begin{table}[!hhh]
\caption{\it  The efficiency defined  by Ê\eqref{eff}   for  four
different estimators based on   $2N$ observations. The drift function is given by  \eqref{m3}
}
\begin{center}
\begin{tabular}{|l|c|c|c|}
\hline
 $~~~~\tilde\theta^{\rule{0mm}{2.5mm}}$ & $N=3$ & $N=5$ & $N=10$ \\
\hline
$\hat\theta_{BLUE } (N,N) $ &0.69608&0.95988&0.99791\\
$\hat\theta_{BLUE} (2N-2,2)$&0.86903&0.99379&0.99981\\
$\hat\theta_{BLUE} (2N,0)$  &0.10040&0.33338&0.62529\\
$\hat\theta_{OLSE} (2N,0)$  &0.08873&0.14103&0.11890\\
\hline
\end{tabular}
\end{center}
\label{tab3}
\end{table}

The results of Table \ref{tab1} -- \ref{tab3} are very  typical for many regression models  with differentiable error processes
(i.e. $q=1$)  and can be summarized as follows. Any   BLUE is far superior to the OLSE and any  BLUE becomes very efficient when
 $N$ is large. Moreover,  the use of information from the  derivatives in constructing BLUEs makes them more efficient
than the BLUE which only uses values of  $\{ y(t) | t \in  \T \} $.
We also emphasize that the BLUEs which
use more than  two values of  the derivative  $y^\prime $  of the  process
 have lower efficiency than the BLUE that uses exactly two values of derivatives, $y^\prime(A)$ and $y^\prime(B)$.
Therefore
  the best way of constructing the BLUE for $N$ observations in the interval $[A,B]$ is to emulate the asymptotic BLUE: that is, to use $y^\prime(A)$ and $y^\prime(B)$ but
  for the other $N-2$ observations use values of  the process $\{ y(t) | t\in \T \} $.

%
%
%

\section{Appendix}
\label{sec:appendix}
\def\theequation{6.\arabic{equation}}
\setcounter{equation}{0}

\subsection{Proof of Lemma \ref{lem:constr-unbiased}}

 The mean of  $\hat\theta^T_{G}$ is
  \bea
\mathbb{E} [\hat\theta^T_{G}]= \theta^T \int_\T F(t) G^T(dt)= \theta^T  \sum_{i=0}^q \int_\T f^{(i)}(t)G^T_i(dt)\, .
\eea
This implies  that the estimator $\hat\theta_{G}$ is unbiased if and only if
\be
\label{eq:unbias-cond}
 \sum_{i=0}^q \int_\T{f^{(i)}}(t) G^T_i(dt)=I_m.
\ee

Since $G_i=C^{-1}\zeta_i$, we have
  \bea
  \sum_{i=0}^q \int_\T{f^{(i)}}(t) G^T_i(dt)=
   \sum_{i=0}^q \int_\T f^{(i)}(t)\zeta^T_i(dt) {C^{-1}}^{T} = C^T {C^{-1}}^{T}=I_m \, ,
\eea
which completes the proof.

%

\subsection{Proof of Theorem \ref{th:GenGMq1}  } $\,$

{\bf I.}
We will call a signed matrix-measure $G$ unbiased if the associated estimator $ \widehat{ \theta}_G $ defined in \eqref{eq:est} is unbiased.
The set  of all unbiased signed matrix-measures will be denoted by $\mathcal{S}$. This set is convex.

The covariance matrix of any  estimator $ \widehat{ \theta}_G$ is the matrix-valued function
$ \phi(G) = {\rm Var} (\hat \theta_G)$ defined in \eqref{eq:var}.
%
The BLUE minimizes this matrix-valued function on the set $\mathcal{S}$.

Introduce the vector-function $d: \mathcal{T} \times \mathcal{S} \to \mathbb{R}^m$  by
$$
 d(s,G)= \sum_{j=0}^q \int_\mathcal{T} K^{(j)}(t,s) G_j(dt) - \phi(G) f(s)\,.
$$
The validity of  \eqref{eq:suff-ness-cond-q} for all $s\in\mathcal{T}$ is equivalent to
the validity of
$d(s,G)=0_{m\times 1}$ for all $s\in\mathcal{T}$.
Hence we are going to prove that $ \widehat{ \theta}_G$ is the BLUE if and only if
$d(s,G)=0_{m\times 1}$ for all $s\in\mathcal{T}$. For this purpose
we will need the following auxiliary result.

\begin{lemma}
\label{lem:unb}
For any  $G \in \mathcal{S}$ we have
$$
 \int_\mathcal{T} \mathbf{d}(s,G)G^T(ds)=0_{m\times m},
$$
where $\mathbf{d}(s,G)=(d(s,G),d^{(1)}(s,G),\ldots,d^{(q)}(s,G))$ is a $m \times (q+1)$ matrix.
\end{lemma}

\textbf{Proof of Lemma~\ref{lem:unb}} Using the unbiasedness condition \eqref{eq:unbias-cond} for $G$, we have
\bea
 \int_\mathcal{T} \mathbf{d}(s,G)G^T(ds)&=&\int_\mathcal{T}\int_\mathcal{T} G(dt)\mathbf{K}(t,s)G^T(ds) - \phi(G)
 \int_\mathcal{T} F(s)G^T(ds) \\
 &=&\phi(G)-\phi(G)I_m=0_{m\times m}
\eea
where $F(s)=(f(s),f^{(1)}(s),\ldots,f^{(q)}(s))$.
\hfill$\Box$

\medskip

For any two  measures $G$ and $H$ in $\mathcal{S}$, denote
$$
 \Phi(G,H)=
 \int_{\mathcal{T}} \int_{\mathcal{T}} G(dt) {\bf K} (t,s) H^T(ds)
$$
which is a matrix of size $m \times m$.
Note that $\phi(G)=\Phi(G,G)$ for any $G \in \mathcal{S}$.

For any two matrix-measures $G$ and $H$ in $\mathcal{S}$ and any real $\alpha$, the  matrix-valued function
$$\phi((1-\alpha)G+\alpha H)=
 (1-\alpha)^2\phi(G)+ \alpha^2 \phi(H)+
  \alpha (1-\alpha)\left[ \Phi(G,H) + \Phi(H,G)
  \right]
$$
is quadratic in~$\alpha$.
Also we have $ \partial^2 \phi((1-\alpha)G+\alpha H)/\partial \alpha^2= 2 \phi(G-H) \geq 0$
and hence  $\phi(\cdot)$ is a matrix-convex function on the space $\mathcal{S}$ (see e.g. \cite{hansen2007differential} for properties of matrix-convex functions).

Since the matrix-function $\phi((1-\alpha)G+\alpha H)$ is quadratic and convex in $\alpha \in \mathbb{R}$, the assertion that
 $G$ is the optimal matrix measure minimizing $\phi(\cdot)$ on $\mathcal{S}$, is equivalent to
  \be
\label{eq:equiv}
\frac{  {\partial} \phi((1-\alpha)G+\alpha H)}  {\partial \alpha}\Big|_{\alpha=0}=0,\;\; \forall \;H \in \mathcal{S}\, .
  \ee

The directional derivative of $\phi((1-\alpha)G+\alpha H)$ as $\alpha\to0$ is
\be
\label{eq:part}
 \frac{\partial}{\partial \alpha}\phi((1-\alpha)G+\alpha H)\Big|_{\alpha=0}=
 \Phi(G,H)+ \Phi(H,G) - 2\phi(G).
 \ee
To rewrite \eqref{eq:part}, we note that $\int_\mathcal{T} \mathbf{d}(s,G)H^T(ds)$ can be written as
 \be
\label{eq:part1}
 \int_\mathcal{T} \mathbf{d}(s,G)H^T(ds)&=&
 \Phi(G,H)- \phi(G) \int_\mathcal{T} F(s)H^T(ds)\\ \nonumber
 &=&\Phi(G,H)- \phi(G),
 \ee
 where in the last equality we have used the unbiasedness condition \eqref{eq:unbias-cond} for $H$.
 Using \eqref{eq:part}, \eqref{eq:part1} and the fact that the matrix $\Phi(H,G)- \phi(G)$ is the transpose of $\Phi(G,H)- \phi(G)$ we obtain
  \be \nonumber
 \frac{\partial}{\partial \alpha}\phi((1\!-\!\alpha)G+\alpha H)\Big|_{\alpha=0}
 &=& \int_\mathcal{T} \mathbf{d}(s,G)H^T(ds)+\left[\int_\mathcal{T} \mathbf{d}(s,G)H^T(ds)\right]^T \,.
 \\
 \label{eq:bb}
 \ee

Consequently, if
$d(s,G)=0_{m\times 1}$ for all $s\in\mathcal{T}$, then
\eqref{eq:equiv} holds and hence $G$ gives the BLUE.

{\bf II.}
Assume now  that   $G$ gives the BLUE $\hat \theta_G$. This implies, first, that  \eqref{eq:equiv} holds and second,
for all $c \in \mathbb{R}^m$  $c^T\phi(G)c\le c^T\phi(H)c$, for any $H\in\mathcal{S}$.
Let us deduce  that  $d(s,G)=0_{m\times 1}$ for all $s\in\mathcal{T}$ (which is equivalent to validity of \eqref{eq:suff-ness-cond-q}).
 We are going to prove this by contradiction.

Assume  that there exists $s_0\in{\cal T}$ such that $d(s_0,G)\neq 0$.
Define the signed matrix-measure $\zeta=(\zeta_0,\zeta_1,\ldots,\zeta_q)$ with $\zeta_0(ds)=G_0(ds) + \kappa d(s_0,G)\delta_{s_0}(ds)$, $\kappa\neq0$, and
$\zeta_i(ds)=G_i(ds)$ for $i=1,\ldots,q$.

Since $G$ is unbiased, $C_G=\int_{\cal T} G(dt) F^T(t)=I_m$. For any small positive or small negative~$\kappa$, the matrix
$C_\zeta=\int_{\cal T} \zeta(dt) F^T(t)=I_m+ \kappa d(s_0,G) f^T(s_0)$ is non-degenerate and its eigenvalues are close to 1.

In view of Lemma~\ref{lem:constr-unbiased}, $H(ds)=C_{\zeta}^{-1}\zeta(ds)$  is an unbiased matrix-measure.
Using the identity  \eqref{eq:bb} and Lemma~\ref{lem:unb} we obtain for the measure $G_\alpha=(1-\alpha)G+\alpha H$:
\bea
\frac{ {\partial} \phi(G_\alpha)} {\partial \alpha}\Big|_{\alpha=0}
 &=& \kappa d(s_0,G)d^T(s_0,G){C_{\zeta}^{-1}}^{T}+\kappa{C_{\zeta}^{-1}}d(s_0,G)d^T(s_0,G).
\eea

Write this as
 $
 {\partial} \phi(G_\alpha)/ {\partial \alpha}\big|_{\alpha=0} = \kappa ( X_0A^T+AX_0)$, where $A={C_{\zeta}^{-1}}$ and $X_0=d(s_0,G)d^T(s_0,G)$ is
 a symmetric matrix.

For any given $A$, the homogeneous Lyapunov matrix equation $XA^T\!+\!AX\!=\!0$
has only the trivial solution $X=0$ if and only if $A$ and $-A$ have no common eigenvalues,
see [\S 3, Ch. 8 in \cite{gantmacher1959matrix}]; this  is the case when $A={C_{\zeta}^{-1}}$ and $\kappa$ is small enough.

This yields that for $X=X_0$, the matrix $X_0A^T+AX_0$ is a non-zero symmetric matrix.
Therefore, there exists a vector $c \in \mathbb{R}^m$ such that  the directional derivative of $c^T\phi(G_\alpha)c$ is non-zero.
 For any such~$c$, $c^T\phi(G_\alpha)c < c^T\phi(G)c  $  for either small positive or small negative $\alpha$
 and hence   $\hat\theta_G$ is not the BLUE.
Thus,  the assumption of the existence of an  $s_0\in{\cal T}$ such that $d(s_0,G)\neq 0$ yields
a contradiction to the fact that $G$ gives the BLUE.
This completes the proof that the equality \eqref{eq:suff-ness-cond-q}
is necessary and sufficient for the estimator $\hat\theta_G$ to be the BLUE .

%
%

\subsection{Proof of Lemma \ref{lem:varG=varH}}
We repeat $i$ times  the  integration by parts formula
 \bea
 \int_\T  \psi^{(i)}(t) {\varphi}(t)dt=\psi^{(i-1)}(t) {\varphi}(t)\Big|_{A}^B
 - \int_\T \psi^{(i-1)}(t) {\varphi}^{(1)}(t)dt \,
\eea
for any differentiable function $\psi(t)$. This gives
 \bea
 \int_\T  \psi^{(i)}(t) {\varphi}(t)dt\!=\!\sum_{j=1}^{i} (-1)^{j-1} \psi^{(i-j)}(t) {\varphi}^{(j-1)}(t)\Big|_{A}^B+(-1)^i\!\int_\T \psi(t){\varphi}^{(i)}(t)dt.
\eea

Using the above equality with $\psi(t)=y^{(i)}(t)$ we obtain that the expectation of two estimators coincide.
Also, using the above equality with $\psi(t)=K^{(i)}(t,s)$ we obtain that the covariance matrices  of the two estimators coincide.

\subsection{Proof of Proposition \ref{prop:CONDIT-MARKOV}}
Straightforward calculus shows that
\bea
 \int_\T K(t,s)\zeta(dt)
 &=&K(A,s)z_A+K(B,s)z_B
 -\int_\T K(t,s)f^{(2)}(t)dt/(\lambda_1+\lambda_2)\\
 &=&K(A,s)z_A+K(B,s)z_B\\
 &&+\big[-K(t,s)f^{(1)}(t)|_A^s+K^{(1)}(t,s)f(t)|_A^{s-}\\
 &&~~~~-K(t,s)f^{(1)}(t)|_s^B+K^{(1)}(t,s)f(t)|_{s+}^B\big]/(\lambda_1+\lambda_2)\\
 &=&(1+\lambda_1A-\lambda_2s)z_A+(1+\lambda_1s-\lambda_2B)z_B +f(s)\\
 &&+\big[K(A,s)f^{(1)}(A)-K^{(1)}(A,s)f(A)\\
 &&~~~ -K(B,s)f^{(1)}(B)+K^{(1)}(B,s)f(B)\big]/(\lambda_1\!+\!\lambda_2)
 \!=\!f(s).
\eea
Therefore, the conditions of Proposition~\ref{th:GenGMq} are fulfilled.

\subsection{Proof of Theorem \ref{th:BLUE-2}}

It is easy to see that $\hat\theta_{G_0,G_1}$  is unbiased.
Further we are going to use Corollary \ref{cor:GGM-1} which
gives the sufficient condition for an estimator to be the BLUE.
We will show that the identity
\be  \label{LHS}
 LHS= \int_{A}^B K(t,s)\zeta_0(dt)+\int_{A}^B K^{(1)}(t,s)\zeta_1(dt) = f(s)
\ee
holds for all $s\in[A,B]$. By the definition of the measure $\zeta$ it follows that
$$
 LHS
  = z_A K(A,s)+z_B K(B,s) + I_A + I_B
 +z_{1,A} K^{(1)}(A,s)+z_{1,B} K^{(1)}(B,s),
$$
where $I_A = \int^s_A K(t,s)z(t)dt, \ I_B=\int^B_s K(t,s)z(t)dt$.
Indeed, for the vector-function $z(t)=\tau_0 f(t)-\tau_2 f^{(2)}(t)+ f^{(4)}(t)$,  we have
\bea
 s_3I_{A}
 &=&\tau_0\!\!\int_A^s K(t,s)  f(t)dt-\tau_2\!\! \int_A^s K(t,s) f^{(2)}(t)dt+ \!\! \int_A^s K(t,s) f^{(4)}(t)dt\\
 &=&\tau_0\!\!\int_A^s K(t,s)  f(t)dt-\tau_2K(t,s)f^{(1)}(t)|_A^s+\tau_2K^{(1)}(t,s)f(t)|_A^s\\
 &&-\tau_2\!\! \int_A^s K^{(2)}(t,s) f(t)dt
 +  K(t,s)f^{(3)}(t)|_A^s-  K^{(1)}(t,s)f^{(2)}(t)|_A^{s-}\\
 &&+ K^{(2)}(t,s)f^{(1)}(t)|_A^{s-}- K^{(3)}(t,s)f(t)|_A^{s-}+ \!\! \int_A^s K^{(4)}(t,s) f(t)dt.
\eea
By construction, the coefficients $\tau_0,\tau_2, $ are chosen such that the equality \eqref{eq:tau4} holds
for all $t\in[A,B]$ and any $s$, implying that integrals in the expression for $I_{A}$ are cancelled.
Thus, we obtain
\bea
 s_3I_{A}
 &=&+\tau_2 K(A,s)f^{(1)}(A)-\tau_2K^{(1)}(A,s)f(A)-  K(A,s)f^{(3)}(A) \\
 && +  K^{(1)}(A,s)f^{(2)}(A)-  K^{(2)}(A,s)f^{(1)}(A)+ K^{(3)}(A,s)f(A)\\
 &&-\tau_2 K(s-,s)f^{(1)}(s)+\tau_2K^{(1)}(s-,s)f(s)+  K(s-,s)f^{(3)}(s) \\
 &&
 - K^{(1)}(s-,s)f^{(2)}(s)+ K^{(2)}(s-,s)f^{(1)}(s)- K^{(3)}(s-,s)f(s).
\eea
Similarly we have
\bea
s_3I_{B}
 &=&-\tau_2K(B,s)f^{(1)}(B)+\tau_2K^{(1)}(B,s)f(B)+  K(B,s)f^{(3)}(B) \\
 && - K^{(1)}(B,s)f^{(2)}(B)+ K^{(2)}(B,s)f^{(1)}(B)- K^{(3)}(B,s)f(B)\\
 &&+\tau_2K(s+,s)f^{(1)}(s)-\tau_2K^{(1)}(s+,s)f(s)-  K(s+,s)f^{(3)}(s) \\
 && + K^{(1)}(s+,s)f^{(2)}(s)- K^{(2)}(s+,s)f^{(1)}(s)+ K^{(3)}(s+,s)f(s).
\eea
Using the assumption on the derivatives of the covariance kernel $K(t,s)$, we obtain
\bea
s_3(I_{A}+I_{B})
 &=& \tau_2 K(A,s)f^{(1)}(A)-\tau_2K^{(1)}(A,s)f(A)-  K(A,s)f^{(3)}(A) \\
 && +  K^{(1)}(A,s)f^{(2)}(A)-  K^{(2)}(A,s)f^{(1)}(A)+ K^{(3)}(A,s)f(A)\\
 &&-\tau_2K(B,s)f^{(1)}(B)+\tau_2K^{(1)}(B,s)f(B)+  K(B,s)f^{(3)}(B) \\
 && -  K^{(1)}(B,s)f^{(2)}(B)+  K^{(2)}(B,s)f^{(1)}(B)- K^{(3)}(B,s)f(B) + s_3f(s).
\eea

Also we have
\bea
 s_3(z_A K(A,s)&\!+\!&z_{1,A} K^{(1)}(A,s))=\\
 &\!=\!&\big(f^{(3)}(A)- \gamma_{1,A} f^{(1)}(A)  + \gamma_{0,A} f(A)\big)K(A,s)\\
 &&+\big(-f^{(2)}(A)+\beta_{1,A}f^{(1)}(A)- \beta_{0,A} f(A)\big)K^{(1)}(A,s)\\
 &\!=\!&f^{(3)}(A)K(A,s)+(-\gamma_{1,A}K(A,s)+\beta_{1,A}K^{(1)}(A,s))f^{(1)}(A)\\
 &&-K^{(1)}(A,s)f^{(2)}(A)+(\gamma_{0,A}K(A,s)-\beta_{0,A}K^{(1)}(A,s))f(A),
\eea
and a similar result at the point $t=B$.
Putting these expressions into \eqref{LHS}
and using the assumption that
constants $\gamma_{1,A},\beta_{1,A},\gamma_{0,A},$ $\beta_{0,A}$ and $\gamma_{1,B},\beta_{1,B},\gamma_{0,B},\beta_{0,B}$
are chosen such that the sum of the functions defined in \eqref{eq:gam-beta-1} is identically equal to zero,
we obtain
$$\int_{A}^B K(t,s)\zeta_0(dt)+\int_{A}^B K^{(1)}(t,s)\zeta_1(dt)= f(s);$$
this completes the proof.

\subsection{Proof of Theorem \ref{th:integr}}
Observing \eqref{26neu} the vector $c$ is can be written as
\bea
 c &=&\int_{a}^A\left[\int_A^B K(t,s)\zeta_0(dt)-  f(s)\right]ds \\
 &=&\int_{a}^A \left[\int_A^B K(t,s')\zeta_0(dt)-  f(s')\right]ds'+\int_{A}^s\left[\int_A^B K(t,s')\zeta_0(dt)-  f(s')\right]ds'\\
  &=&\int_A^B \int_{a}^s K(t,s')ds'\zeta_0(dt)- \int_{a}^s f(s')ds' =
  \int_A^B R^{(1)}(t,s)\zeta_0(dt)- \tilde f(s) .
\eea

We now show that equation \eqref{eq:suff-cond-q} in Proposition~\ref{th:GenGMq} holds for
$K=R, q=1, f= \tilde f$ and $\zeta_i= \tilde  \zeta_i$.
Observing \eqref{thm32cond} and the definition of $\tilde \zeta_i$ in Theorem~\ref{th:integr} we obtain
  \bea
&&  \int_\T  R(t,s) \tilde \zeta_0(dt)+\int_\T  R^{(1)}(t,s) \tilde \zeta_1(dt) \\
&&
 = -  c\left(\int_\T  R(t,s) \eta_0(dt)+\int_\T  R^{(1)}(t,s) \eta_1(dt)\right)+\int_\T  R^{(1)}(t,s) \zeta_0(dt) \\
&& = -  c\cdot 1+\widetilde{f}(s) +c=\widetilde{f}(s) .
\eea

\vspace{ -.5cm }
\subsection{Proof of Propositions \ref{prop:CAR1-expexp} and \ref{prop:CAR1-expcos}}
\label{sec:last}
For the sake of brevity
we only give a proof of Proposition  \ref{prop:CAR1-expexp}, the other result follows by similar arguments.
Direct calculus gives $s_3=K^{(3)}(s+,s)-K^{(3)}(s-,s)=2\lambda_1\lambda_2(\lambda_1+\lambda_2)$.
Then we obtain that the identity \eqref{eq:tau4} holds for $\tau_0=\lambda_1^2\lambda_2^2$ and $\tau_2=\lambda_1^2+\lambda_2^2$.
Straightforward calculations show  that identities \eqref{eq:gam-beta-1} hold with the
specified values of constants $\gamma_1,\gamma_0,\beta_1,\beta_0$.

\bigskip

{\bf Acknowledgements.}
This work has been supported in part by the
Collaborative Research Center ``Statistical modeling of nonlinear
dynamic processes'' (SFB 823, Project C2) of the German Research Foundation (DFG) and the
   National Institute Of General Medical Sciences of the National
Institutes of Health under Award Number R01GM107639. The content is solely the responsibility of the authors and does not necessarily
 represent the official views of the National
Institutes of Health.
The work of Andrey Pepelyshev was partly supported by
the project ``Actual problems of design and analysis for regression models'' (6.38.435.2015)
of St. Petersburg State University.
The authors would like to thank Martina
Stein, who typed parts of this manuscript with considerable
technical expertise.
\bibliography{opt_designs}

\newpage
\begin{frontmatter}
\title{Supplement to ``Best linear unbiased estimators in continuous time regression models''}
\runtitle{Supplement to ``BLUE  in continuous time regression models''}

\begin{aug}
\author{\fnms{Holger} \snm{Dette}\thanksref{m1}\ead[label=e1]{holger.dette@rub.de}},
\author{\fnms{Andrey} \snm{Pepelyshev}\thanksref{m2}\ead[label=e2]{pepelyshevan@cf.ac.uk}}
\and
\author{\fnms{Anatoly} \snm{Zhigljavsky}\thanksref{m2}\ead[label=e3]{zhigljavskyaa@cf.ac.uk}
}
\runauthor{H. Dette et al.}

\affiliation{Ruhr-Universit\"at Bochum\thanksmark{m1}  and Cardiff  University\thanksmark{m2}}

\address{Ruhr-Universit\"at Bochum \\
Fakult\"at f\"ur Mathematik \\
44780 Bochum \\
Germany\\
\printead{e1}}

\address{School of Mathematics\\
 Cardiff  University\\
  Cardiff, CF24 4AG\\
  UK\\
\printead{e2}\\
\printead{e3}}
\end{aug}

\begin{abstract}
We demonstrate that the covariance matrix of
the  BLUE in the continuous-time regression model model with a CAR(2) error process can be obtained as limit of the covariance matrix
of a BLUE in the discrete  regression model with observations at equidistant points and a discrete AR($2$) error process.
\end{abstract}

\begin{keyword}[class=MSC]
\kwd[Primary ]{62K05}
\kwd[; secondary ]{31A10}
\end{keyword}

\begin{keyword}
\kwd{linear regression}
\kwd{correlated observations}
\kwd{signed measures}
\kwd{optimal design}
\kwd{BLUE}
\kwd{AR processes}
\kwd{continuous autoregressive model}
\end{keyword}

\end{frontmatter}



Here we investigate the approximation of the BLUE for continuous-time regression time models with a CAR($2$) error process
(see Section 4 of the paper) by  the
BLUE  in the model
\be
 y(t_i)=\theta^Tf(t_i)+\epsilon(t_i)\, , \quad A \leq t_1  <  t_2 \ldots < t_{N-1} < t_N \leq B~,
 \label{eq:modeldisc}
\ee
where the errors follow a  discrete AR(2) process. This model will be abbreviated
as DAR($2$) throughout this section . The main difficulty to establish the connection between the discrete and continuous  AR($2$)
cased lies in the fact that the discretised
CAR($2$) process follows  an ARMA($2,1$) model rather than the AR($2$), see \cite{he1989embedding}.
To be precise,  assume that the observations in the continuous-time regression  model
\be
 y(t)=\theta^Tf(t)+\epsilon(t)\, , \quad t \in [A,B],
 \label{eq:model}
\ee
are  taken  at $N$  equidistant points  of the form
\be
\label{points}
t_j =A+(j-1)\Delta~,~ (j=1, \ldots, N)
\ee
on the interval $[A,B]$, where  $\Delta=(B-A)/(N-1)$,
 and  that the errors  $\epsilon_1, \ldots , \epsilon_N$ satisfy
the discrete AR(2) equation
\be
\label{AR2}
 \epsilon_j-a_1\epsilon_{j-1}-a_2\epsilon_{j-2}=\varsigma_j,
\ee
where $\varsigma_j$ are  Gaussian  independent identically distributed random variables
with mean $0$ and variance $\sigma^2_\varsigma=\sigma^2(1+a_2)((1-a^2)-a_1^2)/(1-a_2)$.
Here we make a usual assumption  that the equation  \eqref{AR2} defines the AR(2) process for
$j \in \{ \ldots,-2,-1,0,1,2, \ldots\}$ but we only take the values such that $j \in \{ 1,2, \ldots, N\}$.
Let $r_k=\mathbb{E}[\epsilon_j\epsilon_{j+k}] $  denote the autocovariance function of the AR$(2)$  process $\{ \epsilon_1,\ldots,\epsilon_N \}$
and assume without loss of generality that \mbox{$\sigma^2=1$}.

There are in fact three different  forms of the  autocovariance function  (note that we assume throughout $\sigma^2=1$)
of  CAR(2) processes [see e.g. formulas (14)--(16) in \cite{he1989embedding}], which are given by
\be
 \rho_{1}(t)=\frac{\lambda_2}{\lambda_2-\lambda_1} e^{-\lambda_1|t|}-\frac{\lambda_1}{\lambda_2-\lambda_1} e^{-\lambda_2|t|}\, ,
\label{eq:K-eeC}
\ee
 where $\lambda_1\neq\lambda_2$, $\lambda_1>0$, $\lambda_2>0$, by
\be
\rho_{2}(t)=e^{-\lambda |t|}\Big\{\cos(q |t|)+ \frac{\lambda}{q} \sin(q |t|)\Big\}\,,
\label{eq:K-ecosC}
\ee
where $\lambda>0$, $q>0$, and by
\be
\rho_{3}(t)=e^{-\lambda |t|}(1+ \lambda |t|)\, ,
\label{eq:K-elinC}
\ee
where $\lambda>0$.

Also, there are three forms of autocovariance functions of the discrete  AR(2) process  of the form \eqref{AR2}
[see formulas (11)--(13) in \cite{he1989embedding}], which
  are given by
\be
 ~~r_k^{(1)}=    C p_1^k+(1-C) p_2^k, ~~~C=\frac{(1-p_2^2)p_1}{(1-p_2^2)p_1-(1-p_1^2)p_2}\, ,
\label{eq:K-ee}
\ee
where $j\geq 0$, $p_1\neq p_2$, $0<|p_1|,|p_2|<1$; by
\be
r^{(2)}_k=p^k\big(\cos(bk)+C\sin(bk)\big),~~~ C=\cot(b)\frac{1-p^2}{1+p^2} \, ,
\label{eq:K-ecos}
\ee
where $0<p<1$, $0<b<2\pi$ and $b \neq \pi$, and finally by
\be
r^{(3)}_k=p^k\,(1+k C),~~~ C=\frac{1-p^2}{1+p^2}\,,
\label{eq:K-elin}
\ee
where $0<|p|<1$.
Each form of the autocovariance function should be considered individually.
However, we can formulate the following general statement.

\begin{theorem}
\label{th:approx-ar2}
Consider the multi-parameter model \eqref{eq:model} such that the errors follow the  $\mathrm{AR}(2)$ model.
Assume that $f(\cdot)$ is  four times  continuously differentiable.
Define the following constants depending on the form of the autocovariance function $r_k$.
If $r_k$ is of the form \eqref{eq:K-ee}, set
\bea
 \lambda_1 &=&  -\frac{\ln(p_1)}{\Delta},~\lambda_2= -\frac{\ln(p_2)}{\Delta}, \\
 \tau_0 &=& \lambda_1^2\lambda_2^2,~\tau_2=\lambda_1^2+\lambda_2^2,~\beta_1=\lambda_1+\lambda_2,~\beta_0=\lambda_1\lambda_2,\\
 \gamma_1&=& \lambda_1^2+\lambda_1\lambda_2+\lambda_2^2~,\gamma_0=\lambda_1 \lambda_2(\lambda_1+ \lambda_2),~
 s_3=2\lambda_1\lambda_2(\lambda_1+\lambda_2).
\eea
If $r_k$ is of  the form \eqref{eq:K-ecos}, set
\bea
 \lambda &=& -\frac{\ln(p)}{\Delta},~q= -\frac{b}{\Delta}, \\
 \tau_0&=& (\lambda^2+q^2)^2,~\tau_2=2(\lambda^2-q^2),~\beta_1=2\lambda,~\beta_0=\lambda^2+q^2,\\
 \gamma_1&=& 3\lambda^2-q^2~,\gamma_0=2\lambda(\lambda^2+q^2),~
 s_3=4\lambda(\lambda^2+q^2).
\eea
If $r_k$ is of the form \eqref{eq:K-elin}, set
\bea
 \lambda &=& -\frac{\ln(p)}{\Delta},~
 \tau_0=\lambda^4,~\tau_2=2\lambda^2,~\beta_1=2\lambda,~\beta_0=\lambda^2,\\
 \gamma_1&=&3\lambda^2~,\gamma_0=2\lambda^3,~
 s_3=4\lambda^3.
\eea
For large $N$,
the discrete BLUE $\hat \theta_{\BLUE,N}$ based on $N$ observations at the points \eqref{points} can be approximated by
the continuous estimator
{\small
\bea
 \hat\theta_{}\!=\!D^*\big(z_{1,B}y'(B)\!+\!z_{1,A}y'(A)\!+\!z_Ay(A)\!+\!z_By(B)
 \!+\!\int_\T \!\!\!\!z(t)y(t)dt
 \big)
\eea}
where
\bea
 D^*\!=\! \Big(f^{(1)}(B)z_{1,B}^T\!+\!f^{(1)}(A)z_{1,A}^T\!+\!f(A)z_A^T\!+\!f(B)z_B^T
  \!+\!\int_\T f(t)z^T(t)dt\Big)^{-1}.
\eea
Moreover, for this approximation, we have  $D^*=\lim_{N\to\infty} \mathrm{Var}(\hat \theta_{\BLUE,N})$, i.e.
$D^*$ is the limit of the variance of the discrete BLUE as $N\to\infty$.
Here  the quantities
$z(t)$, $z_A$, $z_B$, $z_{1,A}$ and $z_{1,B}$ in the continuous estimator are  defined by
\bea
 z(t) &=& -\big( \tau_2 f^{(2)}(t)-\tau_0 f(t)-f^{(4)}(t)\big)/s_3, 
  \\
 z_{A} &=& \big(f^{(3)}(A) -\gamma_1 f^{(1)}(A)  + \gamma_0 f(A)\big)/s_3, \nonumber \\
 z_{B}&=& \big(-f^{(3)}(B) +\gamma_1 f^{(1)}(B)  + \gamma_0 f(B)\big)/s_3, \nonumber \\
 z_{1,A}&=& \big(-f^{(2)}(A)+ \beta_1f^{(1)}(A)- \beta_0 f(A)\big)/s_3, \nonumber \\
 z_{1,B} &=& \big(f^{(2)}(B)+ \beta_1f^{(1)}(B)+ \beta_0 f(B)\big)/s_3. \nonumber
\eea
\end{theorem}

{\bf Proof.}  It is well known that
the inverse of the covariance matrix $\Sigma =(\mathbb{E}[\epsilon_j\epsilon_{k}] )_{j,k} $  of the discrete AR$(2)$
 process is a  five-diagonal matrix, i.e.
\be
\label{inverseAR2}
 \Sigma^{-1}=\frac{1}{ S }
 \begin{pmatrix}
 k_{11}&k_{12}&k_2&0  &0&0&\ldots\\
 k_{21}&k_{22}&k_1&k_2&0&0&\ldots\\
 k_2 & k_1    &k_0&k_1&k_2&0&\ldots\\
 0&k_2 & k_1    &k_0&k_1&k_2\\
 \vdots&\ddots&\ddots&\ddots&\ddots&\ddots&\ddots\\
 &&0&k_2& k_1    &k_0&k_1&k_2\\
 &&0&0&k_2&k_1&k_{22}&k_{12}\\
 &&0&0&0& k_2&k_{21}&k_{11}\\
 \end{pmatrix}~,
\ee
where the non-vanishing elements are given by $k_0=1+a_1^2+a_2^2$, $k_1=-a_1+a_1a_2$, $k_2=-a_2$,
$k_{11}=1$, $k_{12}=k_{21}=-a_1$, $k_{22}=1+a_1^2$ and
$S=(1+a_1-a_2)(1-a_1-a_2)(1+a_2)/(1-a_2)$.
Using the explicit form \eqref{inverseAR2} for  $\Sigma^{-1}$
we immediately obtain the following result.

\begin{corollary}
\label{cor:2.1}
Consider the linear regression  model \eqref{eq:model} with  observations at $N$  equidistant points \eqref{points} and errors that follow
the discrete AR$(2)$ model  \eqref{AR2}.
Let $h_i$ be the $i$-th column of matrix $H=X^T\Sigma$ and $f_i=f(t_i)$.
Then  the  vectors $h_1,\ldots,h_N$   can be represented explicitly as follows:
\bea
 ~~~h_1 &\!\!=\!\!&\frac{1}{ S }  \left( k_{11} {f_1} + k_{12} {f_2} +k_2 f_3 \right)  \, , \\
 ~~~h_2 &\!\!=\!\!&\frac{1}{ S }  \left( k_{21} {f_1} + k_{22} {f_2} +k_1 f_3+k_2 f_4 \right) \, , \\
 ~~~h_N &\!\!=\!\!&\frac{1}{ S }  \left( k_{11} {f_N} + k_{21} {f_{N-1}} +k_2 f_{N-2} \right)  \, ,\!\!\!\!\!\! \\
 ~~~h_{N-1} &\!\!=\!\!&\frac{1}{ S }  \left( k_{12} {f_N} + k_{22} {f_{N-1}} +k_1 f_{N-2}+k_4 f_{N-3} \right)  \, ,\!\!\!\!\!\! \\
 ~~~h_i&\!\!=\!\!&\frac{1}{ S }  \left(k_2 f_{i-2}+ k_1 f_{i-1}  +k_0 f_{i} + k_1 f_{i+1} + k_2 f_{i+2}\right) \nonumber
  \eea
  for $i=3, \ldots, N-2$.
\end{corollary}

For the approximation of $h_i$, we have
to study the behavior of  the coefficients which depend  on the autocovariance function $r_k$ of the
AR$(2)$ process \eqref{AR2}.

 In the following subsections we will consider the different types of autocovariance functions separately and prove Theorem  \ref{th:approx-ar2} by deriving approximations for the vectors $h_i$.

\medskip

\textbf{Autocovariances of the form \eqref{eq:K-ee}.}
From Corollary \ref{cor:2.1} we obtain that
\bea
 Sh_i&=&-a_2f_{i-2}+(a_1a_2-a_1)f_{i-1}+(1+a_1^2+a_2^2)f_i+(a_1a_2-a_1)f_{i+1}-a_2f_{i+2}\\
 &=&a_2(2f_i-f_{i-2}-f_{i+2})-(a_1a_2-a_1)(2f_i-f_{i-1}-f_{i+1})\\
 &&+(1+a_1^2+a_2^2-2a_2+2a_1a_2-2a_1)f_i\\
 &=&a_2(2f_i-f_{i-2}-f_{i+2})-(a_1a_2-a_1)(2f_i-f_{i-1}-f_{i+1})\\
 &&+(a_1+a_2-1)^2f_i
\eea
for $i=3,4,\ldots,N-2.$
Now consider the case when
the autocovariance structure of the errors has the form \eqref{eq:K-ee} for fixed $N$.
Suppose that the parameters of the autocovariance function \eqref{eq:K-ee} satisfy
$p_1\neq p_2$, $0<p_1,p_2<1$.
We do not discuss the case with negative $p_1$ or negative $p_2$ because discrete AR$(2)$
 processes with such parameters
do not have   continuous real-valued analogues.
From the Yule-Walker equations we obtain that the coefficients $a_1$ and $a_2$ in \eqref{AR2}
are given by
\be
  a_1=r_1\frac{1-r_2}{1-r^2_1},~~~ a_2=\frac{r_2-r^2_1}{1-r^2_1}\, ,
  \label{eq:a1,a2}
\ee
where $r_1=r_1^{(1)}$ and $r_2=r_2^{(1)}$ are defined  by \eqref{eq:K-ee}.
With the notation
$\lambda_1= -\log(p_1) / \Delta$ and $\lambda_2= -\log(p_2) / \Delta$ with $\Delta=(B-A)/N$
we obtain
\be
\label{T0}
 p_1=e^{-\lambda_1\Delta}, ~~~p_2=e^{-\lambda_2\Delta}.
\ee
We will assume that $\lambda_1$ and $\lambda_2$ are fixed but $\Delta$ is small and consider the properties of different quantities as  $\Delta \to 0$.
By a straightforward   Taylor expansion  we obtain the approximations
\be
 a_1=a_1(\Delta)&=&2-(\lambda_1+\lambda_2)\Delta+(\lambda_1^2+\lambda_2^2)\Delta^2/2+O(\Delta^3),\nonumber \\
a_2=a_2(\Delta)&=&-1+(\lambda_1+\lambda_2)\Delta-(\lambda_1+\lambda_2)^2\Delta^2/2+O(\Delta^3), \nonumber\\
 S=S(\Delta)&=&2\lambda_1\lambda_2(\lambda_1+\lambda_2)\Delta^3+O(\Delta^4), \nonumber\\
\label{T4} C=C(\Delta)&=&\frac{\lambda_2}{\lambda_2-\lambda_1}+\frac{1}{6}\lambda_1\lambda_2\frac{\lambda_1+\lambda_2}{\lambda_1-\lambda_2}\Delta^2+O(\Delta^4).
\ee
Consequently  (observing  \eqref{T0} and \eqref{T4}),  for large $N$ the continuous AR(2) process with autocovariances \eqref{eq:K-eeC} can be considered as an
approximation to the discrete AR(2) process with autocovariances
\eqref{eq:K-ee}.

Since $S=O(\Delta^3)$, $a_1=2+O(\Delta)$ and $a_2=-1+O(\Delta)$, 
it follows
\bea
 S\frac{h_i}{\Delta^4}
 &=&f^{(4)}(t_i)-4a_2\frac{1}{\Delta^2}f^{(2)}(t_i)+(a_1a_2-a_1)\frac{1}{\Delta^2}f^{(2)}(t_i)+\frac{1}{\Delta^4}(a_1+a_2-1)^2f_i+O(\Delta)\\
 &=&f^{(4)}(t_i)+\frac{1}{\Delta^2}(a_1a_2-a_1-4a_2)f^{(2)}(t_i)+\frac{1}{\Delta^4}(a_1+a_2-1)^2f_i+O(\Delta)\\
 &=&f^{(4)}(t_i)-(\lambda_1^2+\lambda_2^2)f^{(2)}(t_i)+\lambda_1^2\lambda_2^2f_i+O(\Delta).
\eea
Thus, the vectors $h_i$, $i=3,\ldots,N-2,$ are approximated by the vector-function
\bea
 z(t)=-\frac{1}{s_3 }\big( (\lambda_1^2+\lambda_2^2)f^{(2)}(t)-\lambda_1^2\lambda_2^2 f(t)-f^{(4)}(t)\big),
\eea
where $s_3=2\lambda_1\lambda_2(\lambda_1+\lambda_2)$.
For  the boundary points we obtain
\bea
 Sh_1&=&f_1-a_1f_2-a_2f_3\\
 &=&(-2f_2+f_3+f_1)+(\lambda_1+\lambda_2)(f_2-f_3)\Delta\\
 &&+((-1/2 f_2+1/2 f_3) \lambda_1^2+f_3 \lambda_1 \lambda_2+(-1/2 f_2+1/2 f_3) \lambda_2^2)\Delta^2\\
 &&+((1/6 f_2-1/6 f_3) \lambda_1^3-1/2 f_3 \lambda_1^2 \lambda_2-1/2 f_3 \lambda_1 \lambda_2^2+(1/6 f_2-1/6 f_3) \lambda_2^3)\Delta^3
 +O(\Delta^4)\\
 &=&\big(f^{(2)}(t_2)-(\lambda_1+\lambda_2)f^{(1)}(t_2)+f_3 \lambda_1 \lambda_2\big) \Delta^2+O(\Delta^3)
\eea
and
\bea
 Sh_2&=&-a_1f_1+(1+a_1^2)f_2+(a_1a_2-a_1)f_3-a_2f_4\\
 &=&(-2 f_1+f_4+5 f_2-4 f_3)+(\lambda_1+\lambda_2) (f_1-4 f_2+4 f_3-f_4) \Delta\\
&&+((-1/2 f_1+1/2 f_4-3 f_3+3 f_2) \lambda_1^2+(2 f_2-4 f_3+f_4) \lambda_2 \lambda_1\\
&&~~~ +(-1/2 f_1+1/2 f_4-3 f_3+3 f_2) \lambda_2^2) \Delta^2\\
&&+((1/6 f_1-5/3 f_2+5/3 f_3-1/6 f_4) \lambda_1^3+(-f_2+3 f_3-1/2 f_4) \lambda_2 \lambda_1^2\\
&&~~~ +(-f_2+3 f_3-1/2 f_4) \lambda_2^2 \lambda_1+(1/6 f_1-5/3 f_2+5/3 f_3-1/6 f_4) \lambda_2^3) \Delta^3 +O(\Delta^4)\\
&=&\big(f^{(2)}(t_3)-2f^{(2)}(t_2)+ (\lambda_1+\lambda_2)(3f^{(1)}(t_2)-f^{(1)}(t_1)-f^{(1)}(t_3))-f_3 \lambda_1 \lambda_2 \big)\Delta^2+O(\Delta^3)\\
&=&\big(-f^{(2)}(t_2)+ (\lambda_1+\lambda_2)f^{(1)}(t_2)-f_3 \lambda_1 \lambda_2 \big)\Delta^2+O(\Delta^3)
\eea
Thus, we can see that
$$
{h_1}=- {h_2}+O(1)= z_{1,A}\frac{1}{\Delta}+O(1),
$$
where
\bea
 z_{1,A}=\frac{1}{s_3}\big(-f^{(2)}(A)+ (\lambda_1+\lambda_2)f^{(1)}(A)- \lambda_1 \lambda_2f(A)\big).
\eea
This means that the vectors $h_1$ and $h_2$ at $t_1$ and $t_2$ are large in absolute value and have different signs.
Similarly, we have
$$
h_N =- h_{N-1}+O(1) = z_{1,B}\frac{1}{\Delta}+O(1)
$$
where
\bea
 z_{1,B}=\frac{1}{s_3}\big(f^{(2)}(B)+ (\lambda_1+\lambda_2)f^{(1)}(B)+ \lambda_1 \lambda_2f(B)\big).
\eea

To do a finer approximation, it is necessary to investigate the quantity
\bea
g:=Sh_1+Sh_2,
\eea
which is  of order $O(1)$.
Indeed, we have
\bea
g&=&(3 f_2-3 f_3-f_1+f_4)+(\lambda_2+\lambda_1) (f_1-3 f_2+3 f_3-f_4) \Delta\\
&&+((- f_1+ f_4-5 f_3+5 f_2)/2 (\lambda_1^2+\lambda_2^2)+(2 f_2-3 f_3+f_4) \lambda_2 \lambda_1) \Delta^2\\
&&+(( f_1-9 f_2+9 f_3- f_4)/6 (\lambda_1^3+\lambda_2^3)\\
&&+(-2 f_2+5 f_3- f_4)/2 (\lambda_1^2 \lambda_2+\lambda_1 \lambda_2^2) ) \Delta^3
+O(\Delta^4)\\
&=&f^{(3)}(t_1)\Delta^3+O(\Delta^4)
+(-f^{(1)}(t_1) (\lambda_1^2+\lambda_2^2)-f^{(1)}(t_1) \lambda_2 \lambda_1) \Delta^3 \\
&&+f(t_1)(\lambda_1^2 \lambda_2+\lambda_1 \lambda_2^2)  \Delta^3
+O(\Delta^4)\\
&=& \big(f^{(3)}(t_1) -(\lambda_1^2+\lambda_1\lambda_2+\lambda_2^2) f^{(1)}(t_1)  + \lambda_1 \lambda_2(\lambda_1+ \lambda_2)f(t_1)\big)\Delta^3+O(\Delta^4)
\eea
and, consequently,
\bea
 h_1+h_2=
 \frac{1}{s_3}\big(f^{(3)}(t_1) -(\lambda_1^2+\lambda_1\lambda_2+\lambda_2^2) f^{(1)}(t_1)  + \lambda_1 \lambda_2(\lambda_1+ \lambda_2)f(t_1)\big)+O(\Delta),
\eea
where $s_3=2\lambda_1\lambda_2(\lambda_1+\lambda_2)$.
Therefore, if $\Delta \to 0$, it follows that $h_1+h_2\approx z_A$, where
\bea
 z_{A}=\frac{1}{s_3}\big(f^{(3)}(A) -(\lambda_1^2+\lambda_1\lambda_2+\lambda_2^2) f^{(1)}(A)  + \lambda_1 \lambda_2(\lambda_1+ \lambda_2)f(A)\big).
\eea
Similarly, we obtain
  $h_N+h_{N-1} \approx z_B$ if $\Delta \to 0$, where
\bea
 z_{B}=\frac{1}{s_3}\big(-f^{(3)}(B) +(\lambda_1^2+\lambda_1\lambda_2+\lambda_2^2) f^{(1)}(B)  + \lambda_1 \lambda_2(\lambda_1+ \lambda_2)f(B)\big).
\eea

\medskip

\textbf{Autocovariances of the form \eqref{eq:K-ecos}.}
Consider the autocovariance function of  the form \eqref{eq:K-ecos}, then the coefficients $a_1$ and $a_2$
are given by \eqref{eq:a1,a2} where
$r_1=r_1^{(2)} $ and $r_2=r_2^{(2)} $ are defined by \eqref{eq:K-ecos}.  With the notations
$\lambda= -\log p / \Delta$ and $q=b/ \Delta$ (or equivalently
 $p=e^{-\lambda\Delta}$ and $b=q\Delta$) we obtain by  a Taylor expansion
\bea
 a_1&=&2-2\lambda\Delta+(\lambda^2-q^2)\Delta^2 +O(\Delta^3),\\
 a_2&=&-1+2\lambda\Delta-2\lambda^2\Delta^2+O(\Delta^3),\\
 S&=&4\lambda(\lambda^2+q^2) \Delta^3+O(\Delta^4)
\eea
and
\bea
 C=\frac{\lambda}{q}- \frac{\lambda (\lambda^2+q^2)}{3q} \Delta^2+O(\Delta^4)
\eea
as $\Delta \to 0$.
Similarly, we have
\bea
 S\frac{h_i}{\Delta^4}
 &=&f^{(4)}(t_i)+\frac{1}{\Delta^2}(a_1a_2-a_1-4a_2)f^{(2)}(t_i)+\frac{1}{\Delta^4}(a_1+a_2-1)^2f_i+O(\Delta)\\
 &=&-2(\lambda^2-q^2)f^{(2)}(t_i)+(\lambda^2+q^2)^2 f_i+O(\Delta).
\eea
Thus, the optimal weights $h_i$, $i=3,\ldots,N-2,$ are approximated by the signed density
\bea
 z(t)=-\frac{1}{s_3 }\big( 2(\lambda^2-q^2)f^{(2)}(t)-(\lambda^2+q^2)^2 f(t)-f^{(4)}(t)\big),
\eea
where $s_3=4\lambda(\lambda^2+q^2)$.
Similarly, we obtain that
\bea
{h_1}&=& - {h_2}+O(1)= z_{1,A}\frac{1}{\Delta}+O(1), \\
{h_N}&=& - {h_{N-1}}+O(1)= z_{1,B}\frac{1}{\Delta}+O(1),
\eea
where
\bea
 z_{1,A} &=&\frac{1}{s_3}\big(-f^{(2)}(A)+ 2\lambda f^{(1)}(A)- (\lambda^2+q^2) f(A)\big),
\\
 z_{1,B}&=&\frac{1}{s_3}\big(f^{(2)}(B)+ 2\lambda f^{(1)}(B)+ (\lambda^2+q^2) f(B)\big).
\eea

Calculating $g:=S{h_1}{}+S{h_2}{}$ we have
\bea
 g&=&(3f_2-f_1-3f_3+f_4)+2\lambda(f_1-3f_2+3f_3-f_4)\Delta\\
 &&+((-f_1+7 f_2-8 f_3+2f_4)\lambda^2+q^2(f_1-3 f_2+2f_3))\Delta^2\\
 &&+((-f_1+7 f_2-4 f_3)\lambda q^2+( f_1-15 f_2+24 f_3-4f_4)/3 \lambda^3)\Delta^3+O(\Delta^4)\\
 &=&f^{(3)}(t_1)\Delta^3- (3\lambda^2-q^2)f^{(1)}(t_1)\Delta^3+2\lambda(\lambda^2+q^2)f(t_1)\Delta^3+O(\Delta^4).
\eea

Therefore,   it follows that  $h_1+h_2\approx P_A$ if $\Delta \to 0$, where
\bea
 z_{A}=\frac{1}{s_3}\big(f^{(3)}(A) -(3\lambda^2-q^2) f^{(1)}(A)  + 2\lambda(\lambda^2+q^2) f(A)\big),
\eea
and  $s_3=4\lambda(\lambda^2+q^2)$.
Similarly, we obtain
  the approximation $h_N+h_{N-1}\approx P_B$ if $\Delta \to 0$, where
\bea
 z_{B}=\frac{1}{s_3}\big(-f^{(3)}(B) +(3\lambda^2-q^2) f^{(1)}(B)  + 2\lambda(\lambda^2+q^2) f(B)\big).
\eea

\medskip

\textbf{Autocovariances of the form \eqref{eq:K-elin}.}
For the autocovariance function \eqref{eq:K-elin}
the coefficients $a_1$ and $a_2$
in the AR$(2)$ process are given by \eqref{eq:a1,a2} where
$r_1=r_1^{(3)} $ and $r_2=r_2^{(3)}$ are defined by \eqref{eq:K-elin}.
With the notation  $\lambda =- \log p / \Delta$ (or equivalently  $p=e^{-\lambda\Delta}$)
we obtain the Taylor expansions
\bea
 a_1&=&2-2\lambda\Delta+\lambda^2\Delta^2+O(\Delta^3),\\
 a_2&=&-1+2\lambda\Delta-2\lambda^2\Delta^2+O(\Delta^3),\\
 S&=&4\lambda^3\Delta^3+O(\Delta^4),\\
 C&=&\lambda\Delta-\frac{\lambda^3}{3}\Delta^3+O(\Delta^5)
\eea
as $\Delta \to 0$. Similar calculations as given in the previous paragraphs give
\bea
 S\frac{h_i}{\Delta^4}
 &=&f^{(4)}(t_i)+\frac{1}{\Delta^2}(a_1a_2-a_1-4a_2)f^{(2)}(t_i)+\frac{1}{\Delta^4}(a_1+a_2-1)^2f_i+O(\Delta)\\
 &=&f^{(4)}(t_i)-2\lambda^2f^{(2)}(t_i)+\lambda^4 f_i+O(\Delta).
\eea
Thus, the vectors $h_i$, $i=3,\ldots,N-2,$ are approximated by the signed density
\bea
 z(t)=-\frac{1}{s_3 }\big( 2\lambda^2f^{(2)}(t)-\lambda^4 f(t)-f^{(4)}(t)\big),
\eea
where $s_3=4\lambda^3$. For the remaining vectors $h_1,h_2,h_{N-1}$ and $h_N$  we obtain
\bea
{h_1}&=& - {h_2}+O(1)= z_{1,A}\frac{1}{\Delta}+O(1), \\
{h_N} &=& - {h_{N-1}}+O(1)= z_{1,B}\frac{1}{\Delta}+O(1),
\eea
with
\bea
 z_{1,A} &=&\frac{1}{s_3}\big(-f^{(2)}(A)+ 2\lambda f^{(1)}(A)- \lambda^2 f(A)\big),
 \\
 z_{1,B}&=&\frac{1}{s_3f}\big(f^{(2)}(B)+ 2\lambda f^{(1)}(B)+ \lambda^2 f(B)\big).
\eea
Calculating $g:=S h_1+S h_2$ we have
\bea
 g&=&(3f_2-3f_3-f_1+f_4)+2\lambda(f_1-3f_2+3f_3-f_4)\Delta\\
 &&-\lambda^2(f_1-7f_2+8f_3-2f_4)\Delta^2\\
 &&+1/3\lambda^3(f_1-15f_2+24f_3-4f_4)\Delta^3 +O(\Delta^4)\\
 &=&f^{(3)}(t_1)\Delta^3-3\lambda^2f^{(1)}(t_1)\Delta^3+2\lambda^3f(t_1)\Delta^3 +O(\Delta^4).
\eea
Therefore, if $\Delta \to 0$, it follows that  $h_1+h_2\approx z_A$, where
\bea
 z_{A}=\frac{1}{s_3}\big(f^{(3)}(A) -3\lambda^2 f^{(1)}(A)  + 2\lambda^3 f(A)\big),
\eea
and   $s_3=4\lambda^3$.
Similarly, we obtain
  the approximation $h_N+h_{N-1}\approx z_B$ if $\Delta \to 0$, where
\bea
 z_{B}=\frac{1}{s_3}\big(-f^{(3)}(B) +3\lambda^2 f^{(1)}(B)  + 2\lambda^3 f(B)\big).
\eea

\end{document}